\documentclass[fleqn,usenatbib]{mnras}

\usepackage{newtxtext,newtxmath}

\usepackage[T1]{fontenc}

\DeclareRobustCommand{\VAN}[3]{#2}
\let\VANthebibliography\thebibliography
\def\thebibliography{\DeclareRobustCommand{\VAN}[3]{##3}\VANthebibliography}

\usepackage{graphicx}	

\title[QPOs in tilted disks]{Disk Tearing Leads to Low and High Frequency Quasi Periodic Oscillations in a GRMHD Simulation of a Thin Accretion Disk}

\author[G. Musoke et al.]{
G. Musoke,$^{1}$\thanks{E-mail: g.musoke@uva.nl}
M. Liska,$^{2}$
O. Porth,$^{1}$
Michiel van der Klis,$^{1}$
Adam Ingram $^{3}$
\\
$^{1}$Anton Pannekoek Institute for Astronomy, University of Amsterdam, Science Park 904, 1098 XH Amsterdam, The Netherlands\\
$^{2}$Institute for Theory and Computation, Harvard University, 60 Garden Street, Cambridge, MA 02138, USA\\
$^{3}$Department of Physics, Astrophysics, University of Oxford, Denys Wilkinson Building, Keble Road, Oxford OX1 3RH, UK
}

\date{Accepted XXX. Received YYY; in original form ZZZ}

\pubyear{2022}

\begin{document}
\label{firstpage}
\pagerange{\pageref{firstpage}--\pageref{lastpage}}
\maketitle

\begin{abstract}
Black hole X-ray binaries (BHXRBs) display a wide range of variability phenomena, from long duration spectral state changes to short-term broadband variability and quasi-periodic oscillations (QPOs). A particularly puzzling aspect is the production of QPOs, which -- if properly understood -- could be used as a powerful diagnostic tool of black hole accretion and evolution. In this work we analyse a high resolution three-dimensional general relativistic magnetohydrodynamic simulation of a geometrically thin accretion disk which is tilted by $65^{\circ}$ with respect to the black hole spin axis. We find that the Lense-Thirring torque from the rapidly spinning 10 $M_\odot$ black hole causes several sub-disks to tear off within $\sim 10-20$ gravitational radii. Tearing occurs in cycles on timescales of seconds. During each tearing cycle the inner sub-disk precesses for 1-5 periods before it falls into the black hole. We find a  precession frequency of $\sim 3\rm Hz$, consistent with observed low-frequency QPOs. In addition, we find a high frequency QPO (HFQPO) with centroid frequency of $\sim55$Hz in the power spectra of the mass-weighted radius of the inner disk. This signal is caused by radial epicyclic oscillations of a dense ring of gas at the tearing radius, which strongly suggests a corresponding modulation of the X-ray lightcurve and may thus explain some of the observed HFQPOs.
\end{abstract}




\section{Introduction}
Quasi periodic oscillations (QPOs) appear to be a common characteristic of accreting systems. First discovered in the lightcurves of accreting neutron stars \citep{van-der-Klis_Jansen_van-Paradijs_1985}, QPOs have now been detected in a variety of systems including active galactic nuclei (AGN), low-mass and high-mass X-ray binaries, ultraluminous X-ray sources and tidal disruption events \citep{MucciarelliEtAl2006,GierlinskiEtAl2008,Pasham_Remillard_Fragile_2019,SmithEtAl2021}. QPOs take the form of more or less narrow peaks ($Q\simeq\nu_0/\Delta \nu \sim 2-200$, where $\nu_0$ is the centroid frequency and $\Delta \nu $ is the full width at half maximum around the centroid frequency) in the power density spectra of the lightcurve \citep[e.g.][for a discussion of high coherence QPOs]{BarretEtAl2005}. In the case of AGN, QPOs have been detected throughout the entire electromagnetic spectrum from radio to gamma-rays \citep[e.g.][]{AckermannAjelloEtAl2015,KidgerTakaloEtAl1992,ZhangWang2021}, but they are most prominent in X-ray lightcurves of accreting stellar mass black holes and neutron stars. 

In this work we focus on QPOs in stellar mass black hole X-ray binaries (BHXRBs).  
QPOs in BHXRBs are categorised as either high frequency QPOs (HFQPOs) if their centroid frequency is between $40-450$Hz \citep[e.g.][]{Remillard_Muno_McClintock2003}, and low frequency QPOs (LFQPOs) if their centroid frequency is between $\sim0.1-30$Hz and their properties are generally associated with the spectral state of the source \citep{vanderKlis2006,GierlinskiEtAl2008,Ingram_Motta2019,RemillardEtAl2006}.
While LFQPOs are a common feature of BHXRBs, HFQPOs are rare and have only been detected in a handful of sources e.g. GRO J1655-40 \citep{Remillard_Morgan_McClintock1999,Strohmayer2001,RemillardEtAl2006}, H1743-322 \citep{Homan_Miller_Wijnands2005,RemillardEtAl2006}, XTE J1550-564 \citep{Homan_Wijnands_vanderKlis2001,Miller_Wijnands_Homan2001,Remillard_Sobczak_Muno2002} and GRS 1915$+$105 \citep{Morgan_Remillard_Greiner1997,Strohmayer2001,Belloni_Soleri_Casella2006} with frequencies spanning a few tens of Hz up to $\sim450$Hz. HFQPOs are of particular interest since they are believed to originate very close to the black hole (BH), as supported by their high coherence and short timescales, and can thus be used to probe the innermost region of the accretion flow where strong gravity plays an important role. 
Moreover, the frequencies of HFQPOs can be remarkably stable upon sizeable changes of the X-ray luminosity \citep{Remillard_Muno_McClintock2002,HomanEtAl2005a}. Once the physical mechanism behind both low and high frequency QPOs are sufficiently understood, QPOs have the potential to encode the mass and spin of the black hole \citep[e.g.][]{StuchlikEtAl2006} in addition to constraining the structure of the disk (when mass and spin are known through other methods).

There is currently no consensus on the origin of QPOs in BHXRBs, though numerous models have been proposed to explain the origin of low and high frequency QPOs. These models can be broadly divided into two categories \citep[e.g.][]{van_den_Eijnden_Ingram_Uttley2016} based on the origin of the observed X-ray variability: Geometric models, in which a \textit{constant intrinsic} X-ray flux is modulated by changes in the apparent geometry of the accretion flow in a quasi-periodic manner, or those based on the intrinsic variability of the X-ray emitting plasma. Intrinsic models associate QPOs with changes in the mass accretion rate \citep{Tagger_Pellat1999}, standing shocks/waves \citep{Chakrabarti_Nandi_Debnath2005,Tagger_Pellat1999}, or some form of oscillation or perturbation in the disk (e.g. the adiabatic perturbations in discoseismic models, see  \citealp{Okazaki_Kato_Fukue1987,Nowak_Wagoner1991,Nowak_Wagoner1993,RezzollaEtAl2003,deAvellar_Porth_Younsi2018,DewberryEtAl2020a,Wagoner_Silbergleit_Ortega-Rodriguez2001,ONeill_Reynolds_Miller_Coleman2009}), a nonlinear resonance condition within the accretion disk \citep[e.g. ][]{Abramowicz_Kluzniak_2001,Kluzniak_Abramowicz2002,RemillardEtAl2006,Kato2004,Torok_Kotrlova_Sramkova2011,Torok_Abramowicz_Kluzniak2005,Belloni_Stella2014}, oscillations at the Bardeen-Petterson alignment radius \citep{Fragile_Mathews_Wilson2001}, or a breathing mode \citep{Dexter_Blaes2014}. The observed ratios of simultaneous HFQPOs (typically 3:2) have particularly motivated the idea that HFQPOs are the result of a resonance condition within the disk \citep{Abramowicz_Kluzniak_2001,Kato2004}.

There is increasing evidence for the geometric origin of some QPOs, such as the dependence of LFQPO amplitude on the inclination of the binary orbit \citep{Heil_Uttley_Klein-Wolt2015,Motta_Casella_Henze2015}, phase lag dependence on inclination for type-C LFQPOs \citep{van-den_Eijnden_Uttley_Motta2017} and the modulation of the equivalent width of the iron line with the phase of a type-C QPO detected in H1743-322 \citep{Ingram2016} (also GRS 1915$+$105, see \citealp{Ingram_van_der_Klis_2015,Ingram2016,Nathan_Ingram_Homan2022}). Models based on geometric effects invoke a wide range of mechanisms for QPO production often resulting from the misalignment between the BH spin axis and the binary orbit. One such model is the relativistic precession model (RPM) proposed by \citet{StellaVietri1998,StellaEtAl1999, Fragile_Straub_Blaes2016} in which the warping of space-time results in nodal precession and periastron precession of a test mass in a Keplerian orbit (e.g. an orbiting hotspot in the disk as in \citealp{Schnittman_Bertschinger2004}, see also \citealp{Beheshtipour_Hoormann_Krawczynski2016}) give rise to LFQPOs and HFQPOs respectively. \cite{IngramEtAl2009} (see also \citealp{Ingram_Done2011,Ingram_Done2012}) extended the RPM model from the original framework of test particle orbits in an accretion disk, by applying the model to truncated \citep{Esin_McClintock_1997ApJ} accretion disks. \cite{IngramEtAl2009} proposed that type-C LFQPOs \citep[see][for a classification of BH-XRB QPOs]{Remillard_Sobczak_Muno2002} can be produced by Lense-Thirring precession of the entire hot geometrically thick component (the corona) of the truncated accretion disk. The RPM model predicts not only the frequency, but also the common association of LFQPOs with the hard component ($>6\rm keV$) in the steep powerlaw state \citep{ChurazovEtAl2001,RemillardEtAl2006} and is one of the leading models of QPO production. Several groups have observed Lense-Thirring precession of geometrically thick disks in GRMHD simulations \citep{Fragile_Blaes_Omer2008,Fragile_Anninos2005,Fragile_Blaes_Omer2007, Liska_Hesp_Tchekhovskoy2018, Liska_phase_lag}, however it was unclear how such a precessing disk forms, and, what determines its size and precession frequency. Disk tearing, a process where a smaller precessing disk tears off from a larger non-precessing disk, was postulated as a possible solution and observed in recent SPH (e.g. \citealt{NixonEtAl2012, NealonEtAl2015}) and GRMHD simulations \citep{Liska_2019_codepaper, Liska_Hesp_Tchekhovskoy2021_disc_tearing_BP}.

In this paper we analyse an extremely high resolution GRMHD simulation of a highly tilted geometrically thin accretion disk, which was first presented in \cite{Liska_2019_codepaper}. In this simulation the disk was found to tear and precess, making it an attractive dataset to look for signatures of precession induced QPOs. In Section \ref{sec:num_setup} we describe the numerical setup. In Section \ref{sec:overview} we describe the evolution of the disk and QPO signatures. In Section \ref{sec:discussion} we compare our work to previous numerical studies and observations before concluding in Section \ref{sec:Conclusion}.

\section{Numerical and Physical Setup} \label{sec:num_setup}
In this work we model the evolution of a very thin accretion disk of aspect ratio $h/r = 0.02$ around a rapidly spinning black hole with spin parameter $a = 0.9375$. This scaleheight is maintained throughout the simulation by cooling the disk to its target temperature as described in \cite{Noble_Krolik_Hawley2009}. In the initial conditions, we initialise the velocities as Keplerian and tilt the disk by $65^{\circ}$ with respect to the horizontal equatorial plane of the black hole. The disk has an inner radius of $6.5 r_g$ and outer radius of $76 r_g$. The radial surface density profile scales as $\Sigma \sim r^{-1}$ while the vertical density profile is modeled by a Gaussian profile with a full-width half maximum (FWHM) equal to the local scale-height of the disk. 

The disk is initially threaded with a purely toroidal magnetic field with vector potential given by  $A_{\theta} \propto (\rho-0.0005)r^{2}$. This vector potential is normalised to yield an approximately uniform plasma-$\beta = p_{gas}/ p_{mag} \sim 7$, where $p_{gas}$ and $p_{mag}$ are the gas and magnetic pressures respectively, such that the disk remains gas pressure dominated. Namely, if the disk would become magnetic pressure dominated, it would be unable to achieve the targeted scaleheight. On the other hand, if the magnetic field would be weaker, we would need to run the simulation for much longer to achieve mass inflow equilibrium in the inner $\sim 20 r_g$.

The simulation was conducted using the GPU-accelerated GRMHD code H-AMR \citep{Liska_2019_codepaper}. H-AMR is a massively parallel 3D GRMHD code based on the methods developed in the GRMHD code HARM2D \citep{Gammie2003,Noble_2006}. H-AMR has been extensively modified from these original sources in order to increase the code's speed and robustness through numerous features including adaptive mesh refinement (AMR), local adaptive time-stepping (LAT), and GPU accelerations. H-AMR utilises a staggered grid for constrained transport of magnetic fields as described in \citep{Gardiner_Stone2005} and solves the GRMHD equations of motion in conservative form in arbitrary (fixed) spacetimes. H-AMR uses a finite volume, shock-capturing Godunov-based HLLE scheme, with second order spatial accuracy (PPM reconstruction of primitive variables, \citealp{Colella_Woodward1984}), and second order accurate time evolution. 

The simulation was performed on a logarithmic spherical-polar grid in a Kerr-Schild foliation and uses outflow boundary conditions in the radial direction, transmissive boundary conditions in the $\theta$-direction and periodic boundary conditions in the $\phi$-direction. The inner boundary was placed just inside the event horizon at $\sim r_g$ and the outer boundary at $10^5 r_g$. The simulation was conducted with an extremely high resolution grid, featuring a maximum effective resolution $N_r \times N_{\theta} \times N_{\phi} = 13440\times4608\times8192$ in the disk beyond $\sim10 r_g$. The grid resolution gradually drops to the base resolution of $1728\times576\times1024$ at $r \sim 1.2r_g$ in order to increase the timestep by an order of magnitude \citep{Liska_2019_codepaper}. The grid uses 3 levels of adaptive mesh refinement (AMR) and 5 levels of LAT. In addition, to prevent the timestep becoming limited by the Courant condition \citep{CourantFriedrichs1928} in $\phi$, we reduce the azimuthal resolution progressively from $N_{\phi}=1024$ to $N_{\phi}=16$ within $30^{\circ}$ of each pole. 

\begin{figure}
    \centering
    \includegraphics[clip,trim=0.0cm 0.0cm 0.0cm 0.0cm,width=0.45\textwidth]{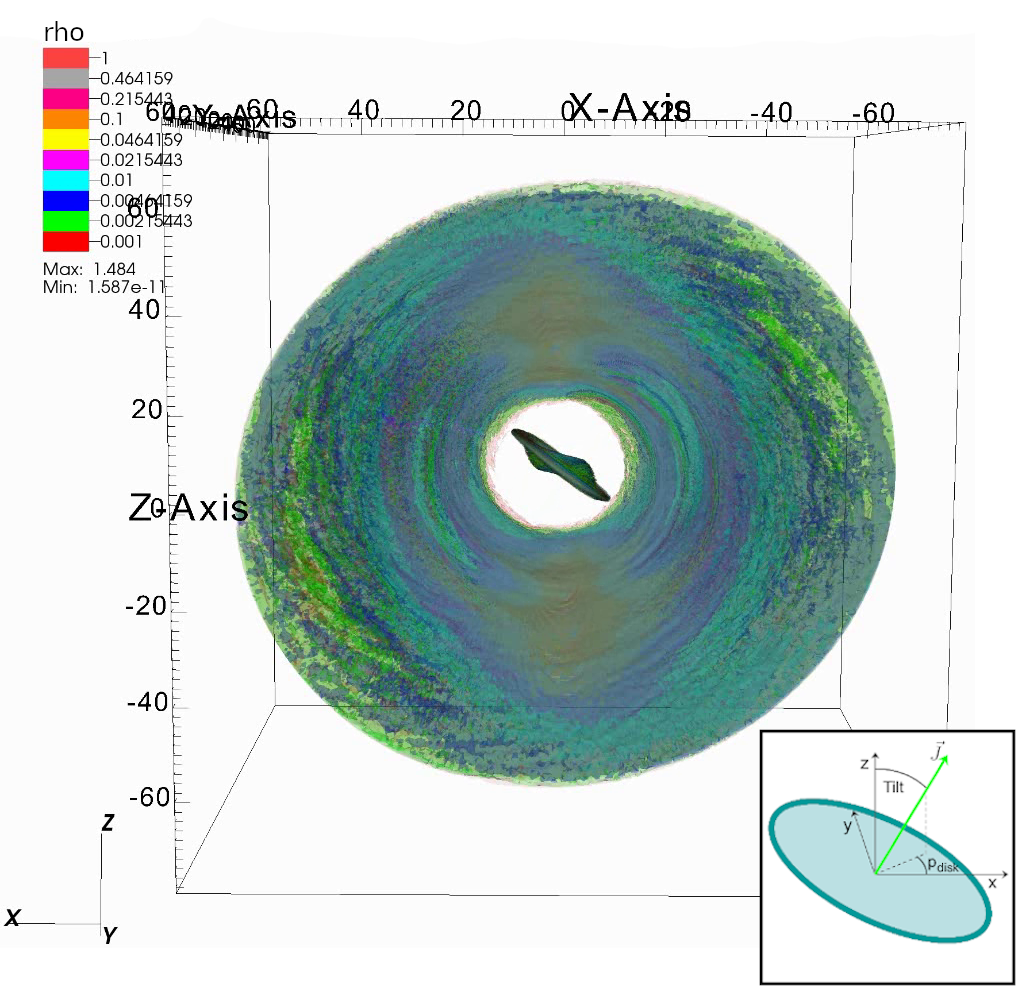}
    \caption{Isocontours of the density during a large tearing event showing a rapidly precessing sub-disk with tearing radius of $\simeq13 \rm r_{\rm g}$.  The inner parts of the sub-disk undergo Bardeen-Petterson alignment which creates a strong warp.  Several dense rings are visible in the outer disk along with an $m=2$ pattern in density .  The inset illustrates the local definition of Tilt- and precession angle ($p_{\rm disk}$) through the angular momentum of a disk annulus. The black hole spin is aligned with the $z$-axis.
    }
    \label{fig:3D}
\end{figure}

\begin{table*}
    \caption{Table showing the start $T_{s}$ and end $T_{f}$ time of the disk tearing events in the simulation, alongside the duration of each tearing event. The start time $T_{s}$ corresponds to the time that the inner disk tears off from the outer disk. $T_{f}$ is the time at which the inner sub-disk is no longer present, having been completely accreted onto the BH. In the fifth column the tearing radii are measured at the time halfway through each tearing event $t = T_{mid} = (T_{s} +T_{f})/2$. The final column lists the tearing radii measured at the start of the tearing cycle (at $t = T_{s}$). We find signatures of HFQPOs in tearing cycles E7 and E10. We note that the simulation ends while the final tearing cycle E10 is still ongoing, and so the duration and tearing radius measured at $t = T_{mid}$ (fifth column) are not representative of the full tearing cycle E10.} \label{tab:disk_tearing}
    \centering
    \begin{tabular}{c|c|c|c|c|c}
        Event & $T_{s} (r_g/c)$ & $T_{f} (r_g/c)$ & Duration $(r_g/c)$ & Tearing radius at $t = T_{mid}$ $(r_g)$ & Tearing radius at $t = T_{s}$ $(r_g)$\\
        \hline
         E1        & 1216     & 2027   & 811    &  7.8    & 13 \\
         E2        & 5876     & 8852   & 2976   &  10     & 14 \\
         E3        & 13618    & 18839  & 5221   &  11     & 18 \\
         E4        & 25903    & 27152  & 1249   &  10     & 13 \\
         E5        & 31715    & 33013  & 1298   &   9     & 13 \\
         E6        & 39922    & 40827  & 905    &  7.5    & 13 \\
         E7        & 47138    & 89420  & 42282  &  14.6   & 20 \\
         E8        & 93878    & 95089  & 1211   &  7.4    & 10 \\
         E9        & 96290    & 101800 & 5510   &  10.5   & 13 \\
         E10       & 113318   & 140183 & 26865  &  13     & 19
    \end{tabular}
\end{table*}

\section{Simulation overview} \label{sec:overview}
\begin{figure*}
    \centering
    \includegraphics[clip,trim=1.0cm 1.8cm 1.0cm 2.3cm,width=14.4cm]{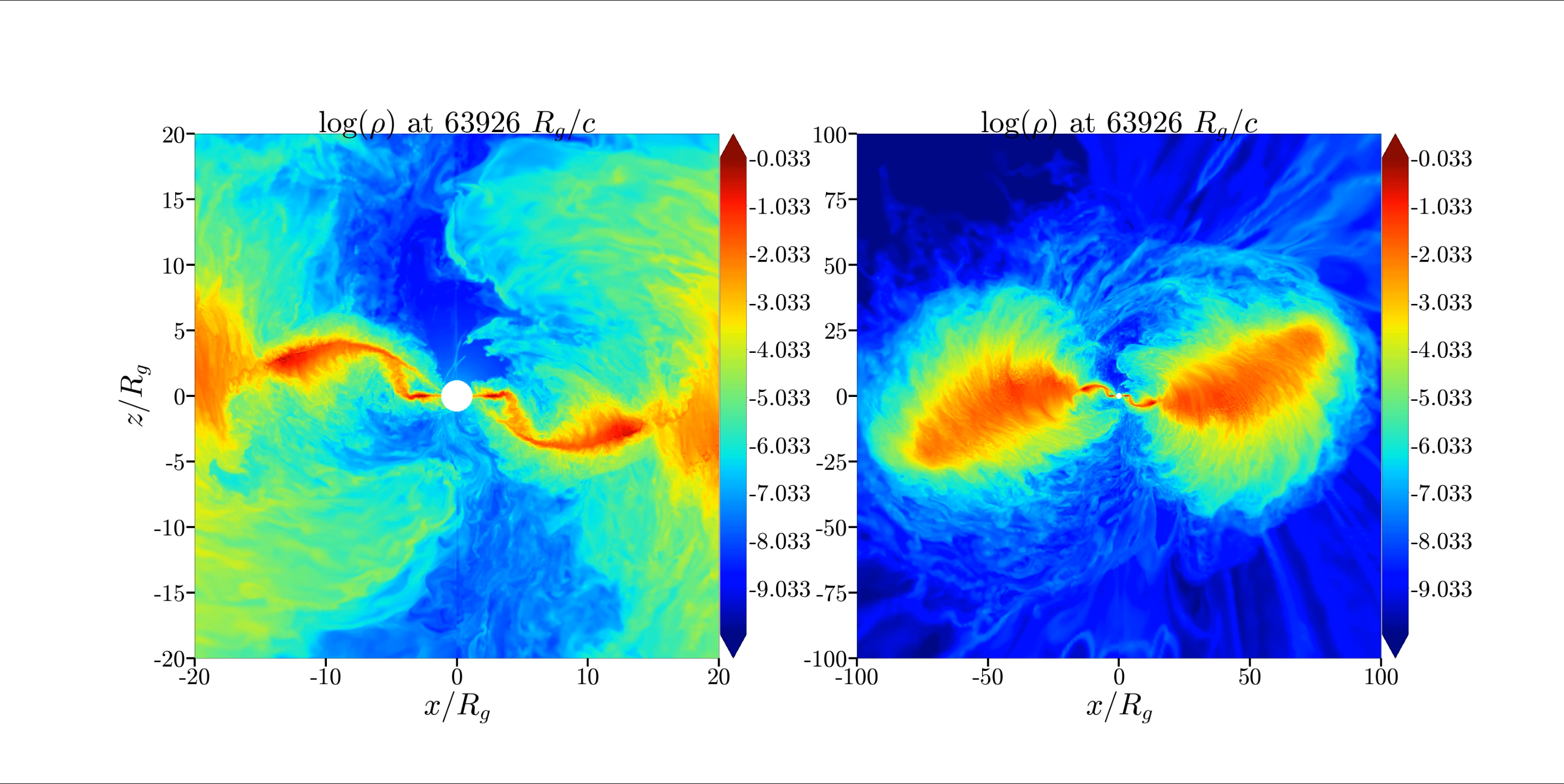}
    \includegraphics[clip,trim=1.0cm 1.8cm 1.0cm 2.3cm,width=14.4cm]{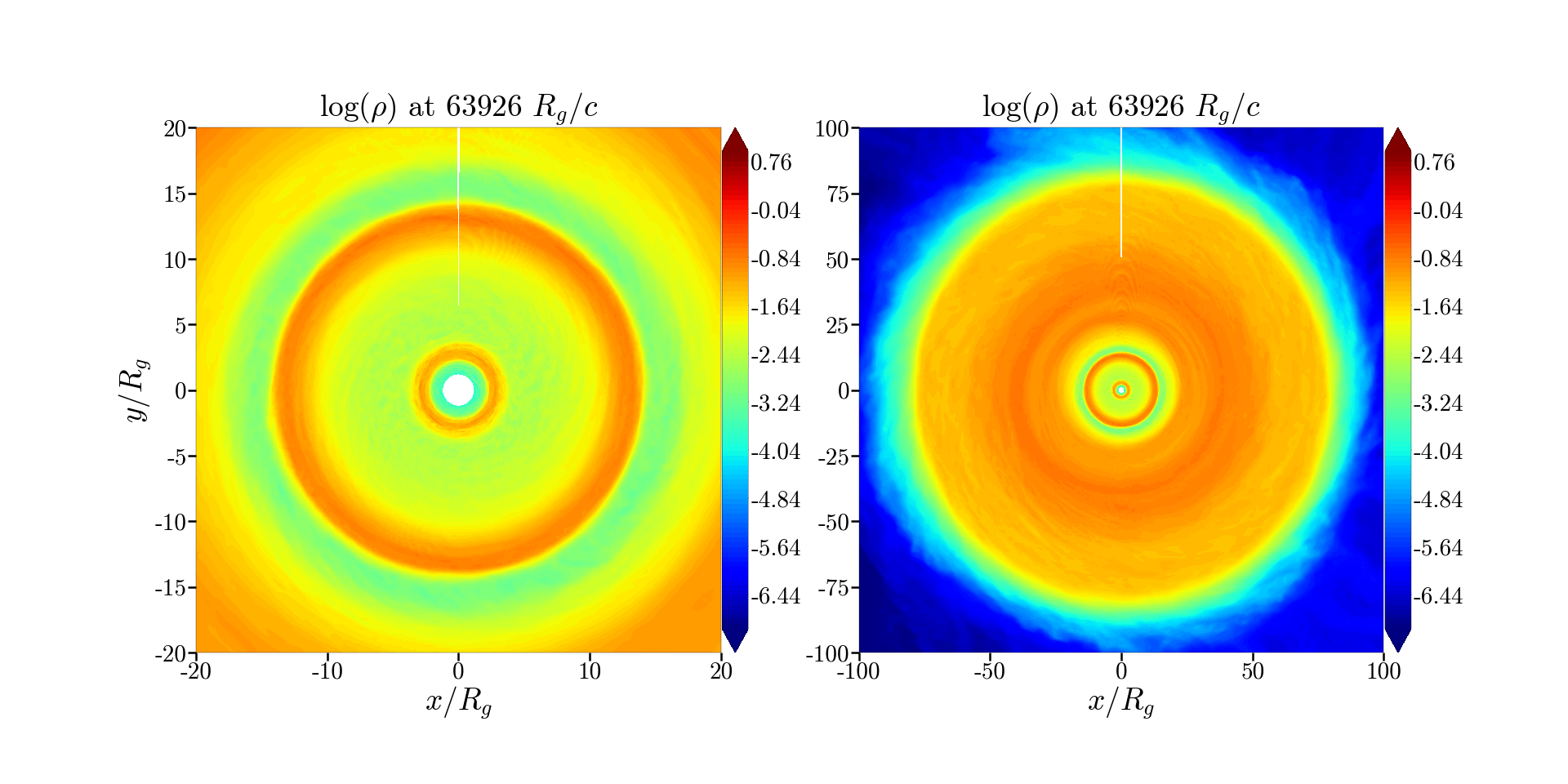}
    \caption{2d views of the simulation, the left panels show a zoomed in view of the right panels. \textbf{Top:} 2D slice through $\phi = 0$ during the tearing event E7 listed in Table \ref{tab:disk_tearing}. The regions of the inner disk closest to the BH have undergone Bardeen-Petterson alignment. The inner disk and outer disk are connected by plunging streams of accreting material. \textbf{Bottom:} Top-down view of the disk showing $\theta$-integrated density $\rho$. These figures have been created by re-orientating each disk annulus according to its local angular momentum direction. A prominent dense ring is seen at the outer edge of the torn inner-sub disks at $\sim 13 r_g$. Multiple dense rings are seen at different radii in the disk, disk tearing always occurs behind (i.e. at larger radii) a dense ring.}
    \label{fig:2d_rho}
\end{figure*}

The disk tears at places where the viscous stresses cannot provide rapid enough angular momentum transport to counteract the (effective) differential Lense-Thirring torque from the warping of space-time by the spinning black hole \citep[e.g.][]{NixonEtAl2012}. In our simulation, disk tearing generally results in one precessing sub-disk which is decoupled from the outer disk and precesses independently for $\sim 1-5$ precession cycles before aligning and/or being accreted by the black hole \citep{Liska_2019_codepaper}. A total of 10 such tearing cycles occur in the simulation and their associated times, duration of the tearing cycle and radius at which the disk tears are give in Table \ref{tab:disk_tearing}. To get a sense of the geometry, a 3D rendering of simulation is shown during the longest tearing cycle (E7) in Figure \ref{fig:3D} and a corresponding movie can be inspected on the following \href{https://www.youtube.com/playlist?list=PLDO1oeU33GwlE5_hWjOq0FgDx6I_z856Y}{YouTube channel}. Tearing cycle E7 is further illustrated in Figure \ref{fig:2d_rho} which shows a transverse slice of the vertically integrated density in a coordinate system aligned with the local angular momentum vector.  

Throughout the whole simulation, no large scale poloidal magnetic flux is generated, which leads to the absence of any relativistic jets \citep{BlandfordZnajek1977}. This suggests that our simulation is applicable to the jet-less high-soft state of BHXRBs. 
In previous work, it was demonstrated that geometrically thick accretion disks are able to convert toroidal magnetic flux into poloidal magnetic flux on a $10^4-10^5 r_g/c$ timescale, advect it inwards from $r \sim 10^2 r_g$, and launch powerful jets \citep{Liska_Tchekhovskoy_Quataert2020_magnetic-field_dynamo}. It is unclear if either of these two processes can occur in a geometrically thin accretion disk (e.g. \citealt{Lubow_Papaloizou_Pringle1994}). Since the viscous time for thin disks is $\sim 10^2$ times longer than for the much thicker disk presented in \citet{Liska_Tchekhovskoy_Quataert2020_magnetic-field_dynamo}, answering such questions will require much longer GRMHD simulations than presented in this work.

\subsection{Disk Tearing}
\label{sec:tearing}
\begin{figure*}
    \centering
    \includegraphics[clip,trim=1.0cm 1.5cm 1.0cm 2.0cm,width=8.3cm]{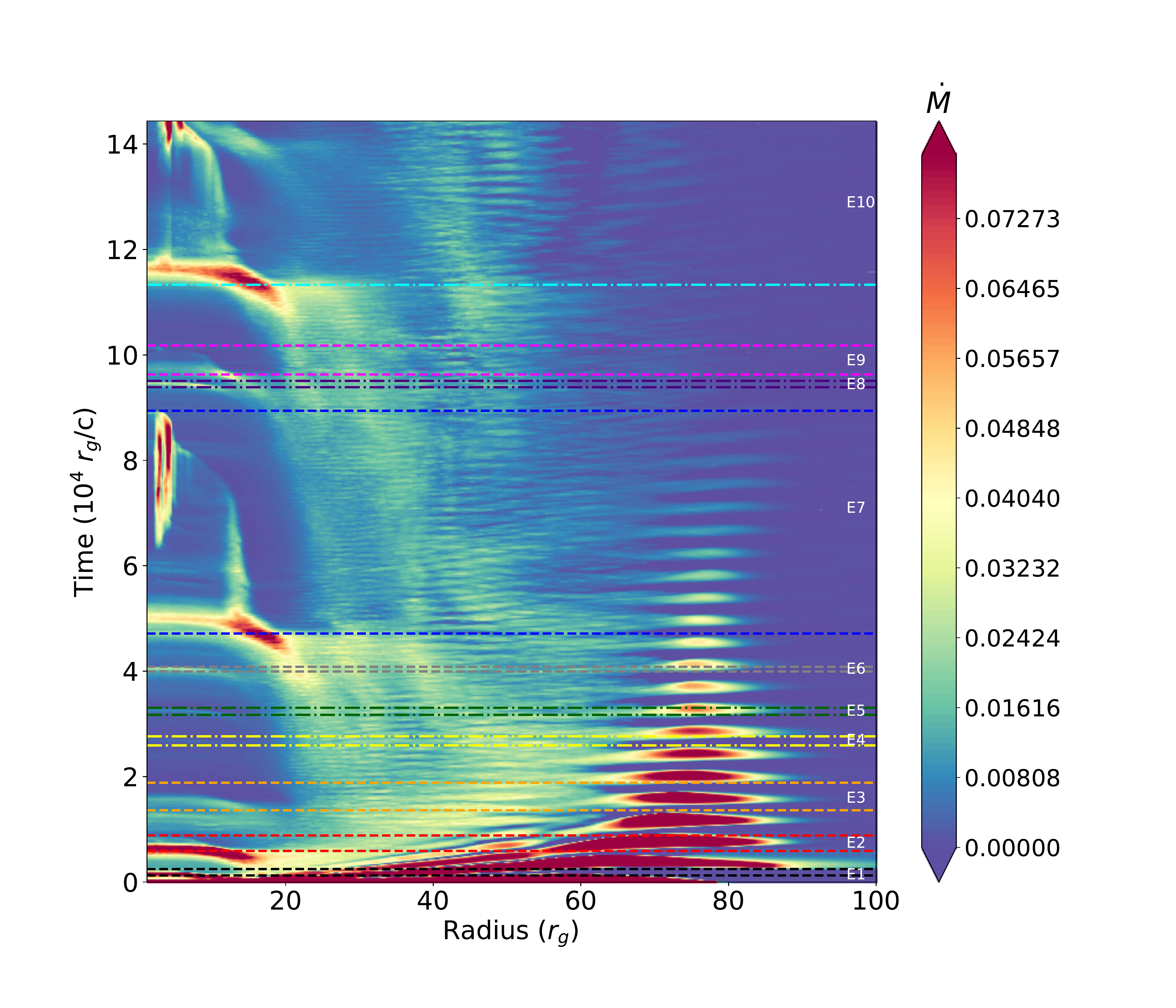}
    \includegraphics[clip,trim=1.0cm 1.5cm 1.0cm 2.0cm,width=8.3cm]{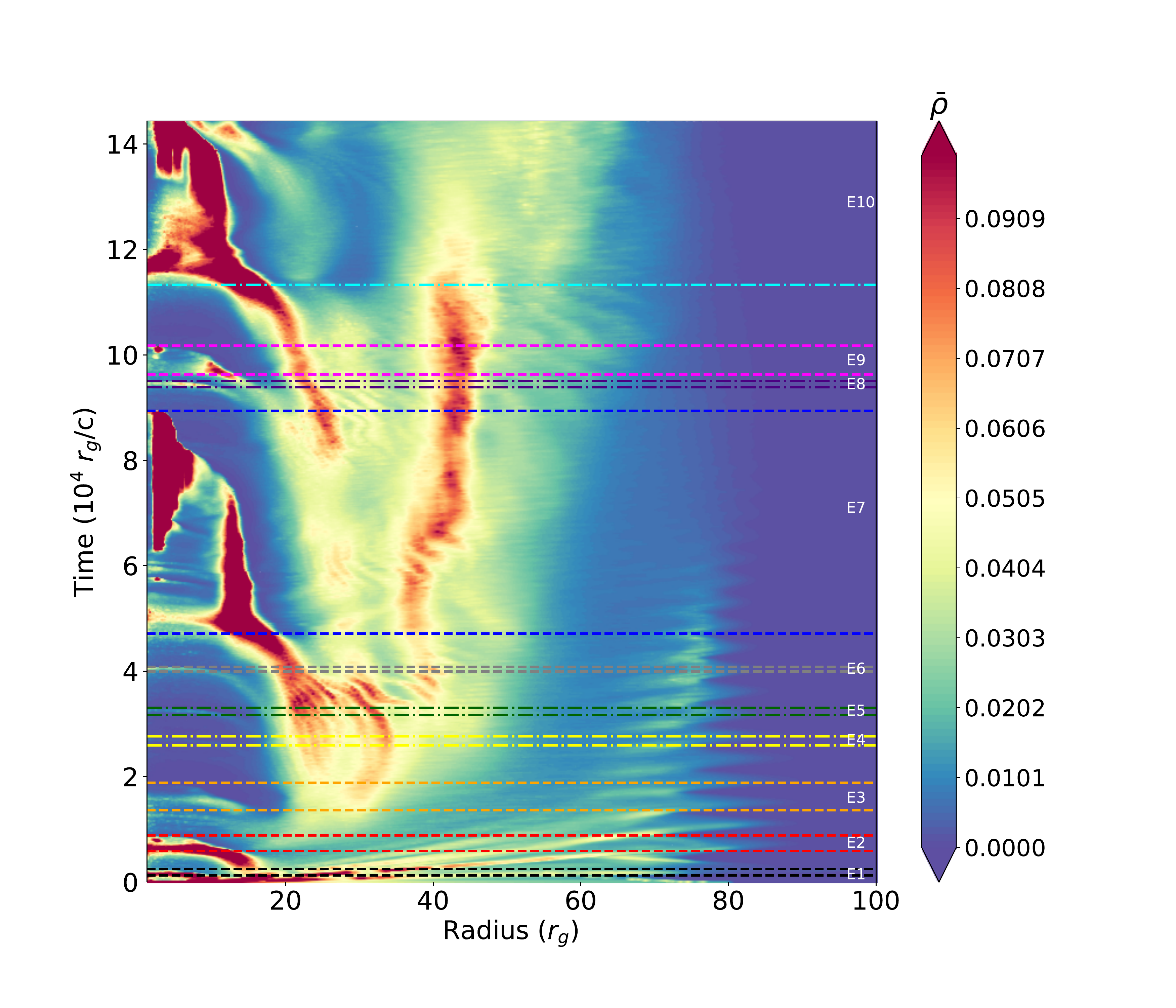}
    \includegraphics[clip,trim=1.0cm 1.5cm 1.0cm 2.0cm,width=8.3cm]{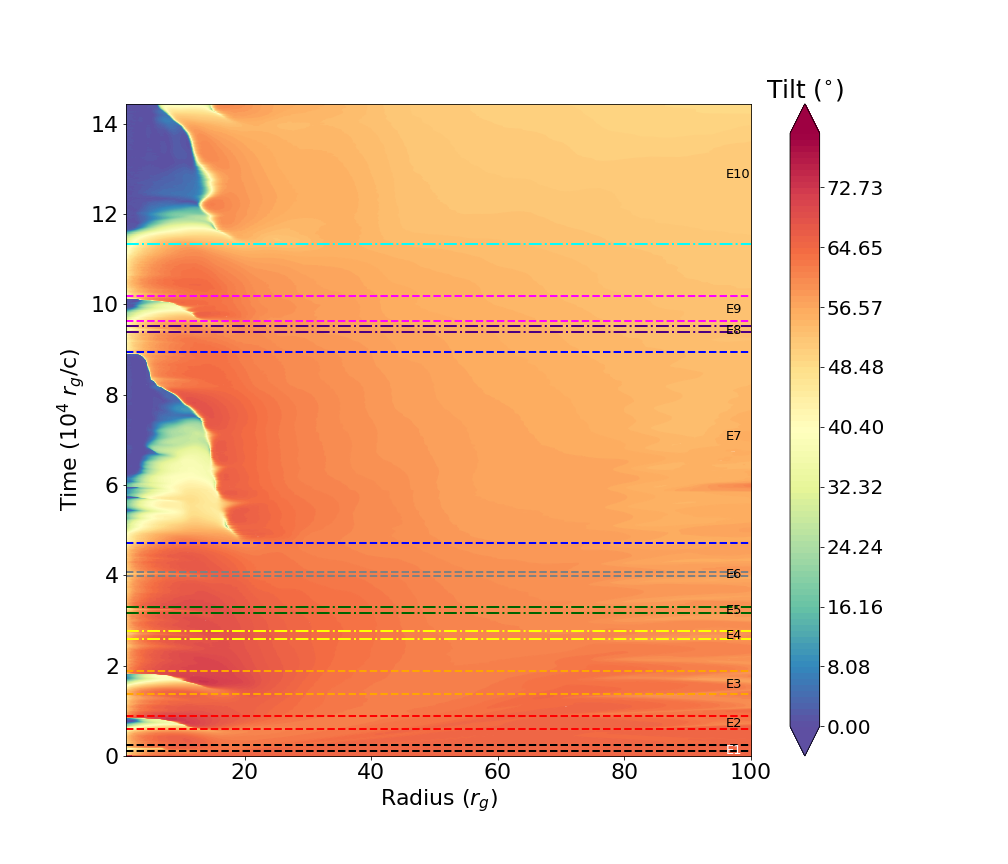}
    \includegraphics[clip,trim=1.0cm 1.5cm 1.0cm 2.0cm,width=8.3cm]{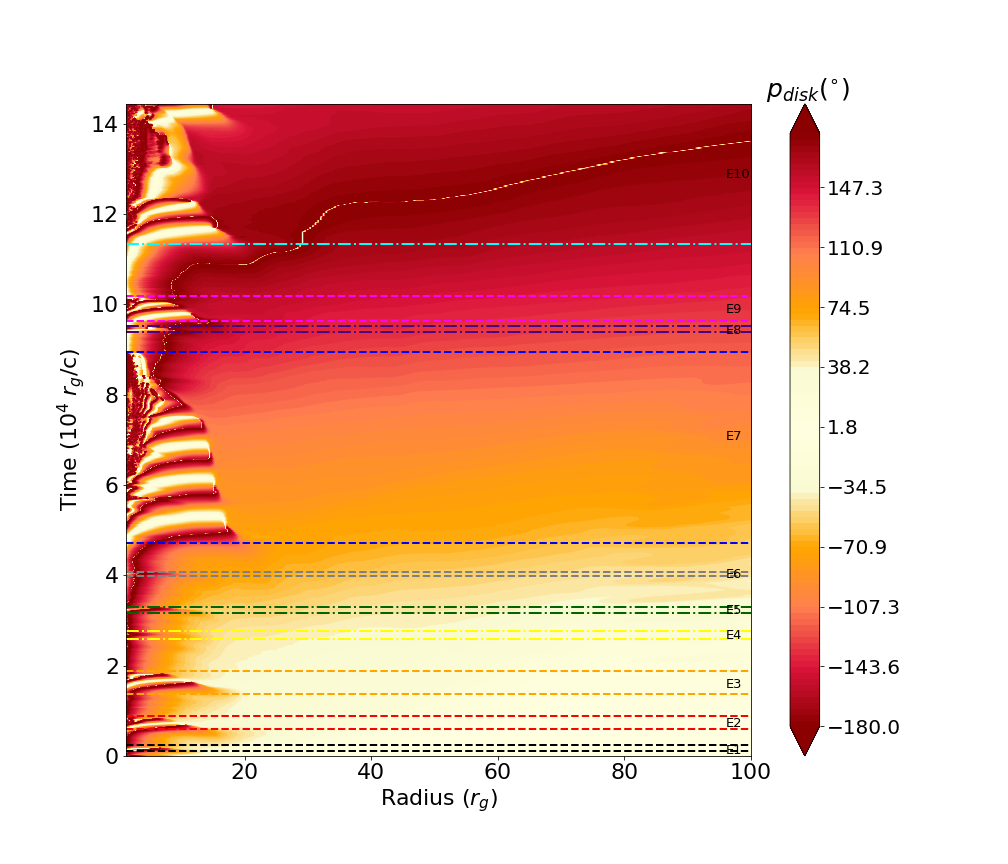}
    \caption{Space-time diagram for the radial mass flux $\dot{M}$ (\textbf{top left}, see equation \ref{eq:Mdot}), average density $\bar{\rho}$ ($=\int \rho^2 dV/\int \rho dV$, \textbf{top right)}, tilt angle (\textbf{bottom left}), disk precession angle $p_{disk}$ (\textbf{bottom right}). The horizontal lines mark the start and end times for each of the tearing events listed in Table \ref{tab:disk_tearing}. Strong tearing events (e.g. E7 and E10) are rapidly followed by bursts of increased mass flux to the central black hole. This can be seen in the $\dot{M}$ and $\bar{\rho}$ panels. The evolution of the tear is visible as sharp discontinuity of the tilt angle in the bottom left panel. Nearly five full precession periods are observed with a period of $\approx 8\, 000r_g/c$ in the disk precession angle. Scaled to a $10 M_\odot$ black hole, corresponding to a frequency of $\approx 3\rm Hz$. This frequency is in the range of those observed for type-C LFQPOs.
    \label{fig:spacetime}}
\end{figure*}

To better understand the disk tearing process we plot in Figure~\ref{fig:spacetime} the space-time diagrams of the radial mass flux $\dot{M}$,
\begin{equation}
  \dot{M} =  -\int_{0}^{2\pi} \int_{0}^{\pi} \rho u^r \sqrt{-g} \,d\theta \, d\phi \text{ ,}
  \label{eq:Mdot}
\end{equation}
where $\rho$ is the mass density, $u^{r}$ is the radial component of the four-velocity and $g$ is the determinant of the four-metric, the average density given by 
\begin{equation}
\bar{\rho} =\frac{\int \rho^2 dV}{\int \rho dV} \text{ ,}    
\end{equation}
and the tilt angle $Tilt$, and precession angle $p_{\rm disk}$. Calculation of the disk tilt and precession angle is outlined in \citealt{Fragile_Anninos2005, Liska_Tchekhovskoy_Ingram2019_BP2} and the geometry is sketched in the inset of Figure \ref{fig:3D}. The disk tear is identifiable as a sharp discontinuity in precession and tilt angles. All events are associated with dense rings that initially form within $\approx 30r_g$. As the material moves inwards through the strongly warped regions of the disk, the inner ring also moves inwards and a tear develops when the outer edge of the ring is at e.g. $r\simeq 19-20 r_g$ for the longest duration tearing cycles E7 and E10. Subsequently, the inner sub-disk starts precessing independently of the outer disk. This tearing process goes along with a burst in accretion onto the black hole around $t \sim 50,000 r_g/c$ for tearing event E7 and around $t \sim 115,000 r_g/c$ for tearing event E10. The recurrence-time of these tearing events is set by the ring formation time at larger distances.

In addition to the inward-moving rings that are associated with tearing events, we observe one other prominent long-lived feature at $\sim 40 r_g$ in the space-time diagram of the average density, which corresponds to the outer dense ring in the lower right panel of Figure \ref{fig:2d_rho}. Naively, one might think this feature will develop into a full tear on a timescale much longer than the runtime of this simulation. However, since this feature almost disappears at the onset of tearing event E10, this is an unlikely scenario.

We speculate that the density enhancements found in rings are caused by the enhanced energy dissipation and angular momentum transport in the strongly warped regions just outside of these rings. As illustrated by the profiles of $\dot{M}(r,t)$ and $\bar{\rho}(r,t)$, this causes gas to be fed from the adjacent regions leading to a sharp drop in density behind the torn off inner disk.

Material progresses from one torn off sub-disk to the other via streamers that, depending on the relative phase angles of the rings, can provide efficient means of angular momentum cancellation. This shortens the accretion time of the inner disk and prevents the inner sub-disk from viscously spreading, which was shown to inhibit disk precession in previous work \citep{Liska_Hesp_Tchekhovskoy2018}. Interestingly, as can be seen in the averaged density of Figure~\ref{fig:spacetime}, angular momentum cancellation does not cause the inner sub-disk to shrink in radial size immediately. Instead, it first depletes the ring formed at the tearing radius of gas. This allows the inner sub-disk to precess with a semi-constant frequency before it is consumed by the black hole.

The space-time diagrams of radial mass flux and density also demonstrate a strong oscillatory signal in the outer regions which we attribute to the initial conditions: as the disk warp propagates outwards, and the outer edge of the disk establishes radial force balance, the outer disk picks up oscillations. After an adjustment time of $\approx 20\, 000 r_g/c$ these oscillations disappear within a radius of $50 r_g$ before fully disappearing after $\approx 90\, 000 r_g/c$. Similar breathing mode oscillations, eventually attributed to the initial conditions, were observed in other work \citep{Mishra_Kluzniak_Fragile2019}.

\subsection{Disk Alignment}
\label{sec:alignment}
\begin{figure}
    \centering
    \includegraphics[clip,trim=0.0cm 0.0cm 0.0cm 0.0cm,width=8.3cm]{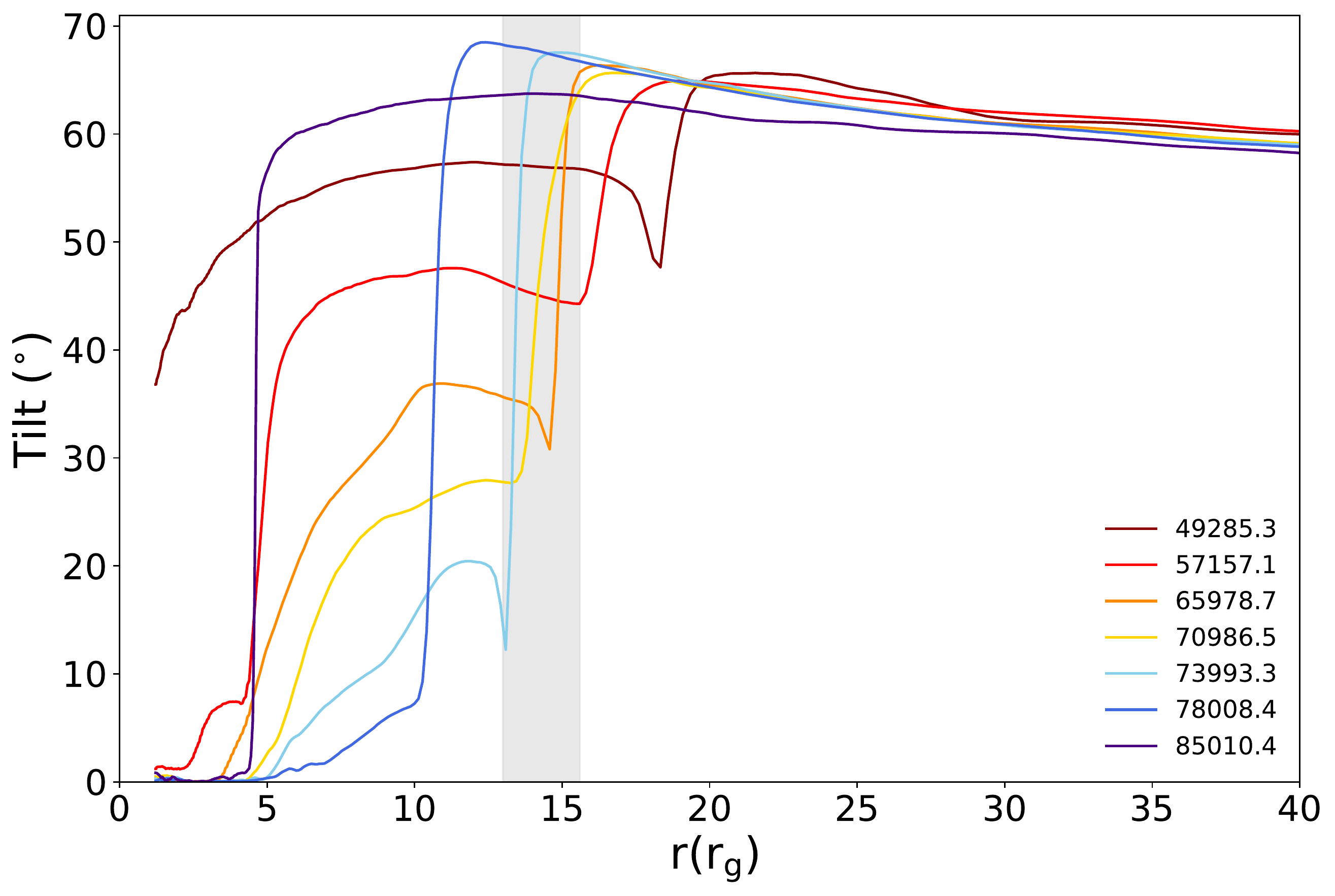}
    \caption{Disk tilt as a function of radius at different evolution times during the tearing cycle E7. Large changes in the tilt angle can be used to identify the location of disk tearing. The tearing radius moves inwards, towards the BH, over the course of the tearing cycle. During the period in which the HFQPO signatures are present, the tearing radius remains fixed between 13 and 15.6 $r_g$ (this range is shown by the solid grey vertical band).}
    \label{fig:tilt_v_rad}
\end{figure}

We observe alignment of the disk with the black hole spin axis during all tearing events. To understand the alignment process, we plot in Figure \ref{fig:tilt_v_rad} the tilt as function of radius for various snapshots during tearing cycle E7. In this figure we can discern two physical process leading to alignment of the disk. 

The first process is the classical \cite{bardeenLenseThirringEffectAccretion1975} (hereafter: BP) effect which forces the inner regions of the disk to align with the black hole spin axis\footnote{In fact, this simulation comprised the first demonstration of Bardeen-Petterson alignment in the absence of a large-scale poloidal magnetic field \citep{Liska_2019_codepaper}.}. BP alignment occurs inside-out within the inner $r \sim 5 r_g$ of the disk and leads to almost full alignment between disk and black hole spin. Interestingly, BP alignment is accompanied by a density increase in the aligned portion of the disk as can be seen in E7, E9 and E10. We speculate that the absence of a warp in the aligned portion of the disk reduces the inflow speed of the gas, which increases the density.

In addition to the BP effect, the disk also exhibits a global alignment mode where the disk aligns outside-in with the black hole spin axis on a timescale similar to the accretion time of the inner sub-disk. This global alignment mode is driven by angular momentum mixing in warps and when streamers deposit misaligned gas from the outer disk into the inner (precessing) sub-disk, which leads to the cancellation of perpendicular angular momentum and causes alignment \citep{Liska_Tchekhovskoy_Ingram2019_BP2, Liska_Hesp_Tchekhovskoy2021_disc_tearing_BP}. We will quantify the relative contribution of these two processes in future work.

\subsection{Detection of a HFQPO at the Tearing Radius} \label{sec:HFQPO_detection}
\begin{figure}
    \centering
    \includegraphics[clip,trim=0.0cm 0.0cm 0.0cm 0.0cm,width=7.0cm]{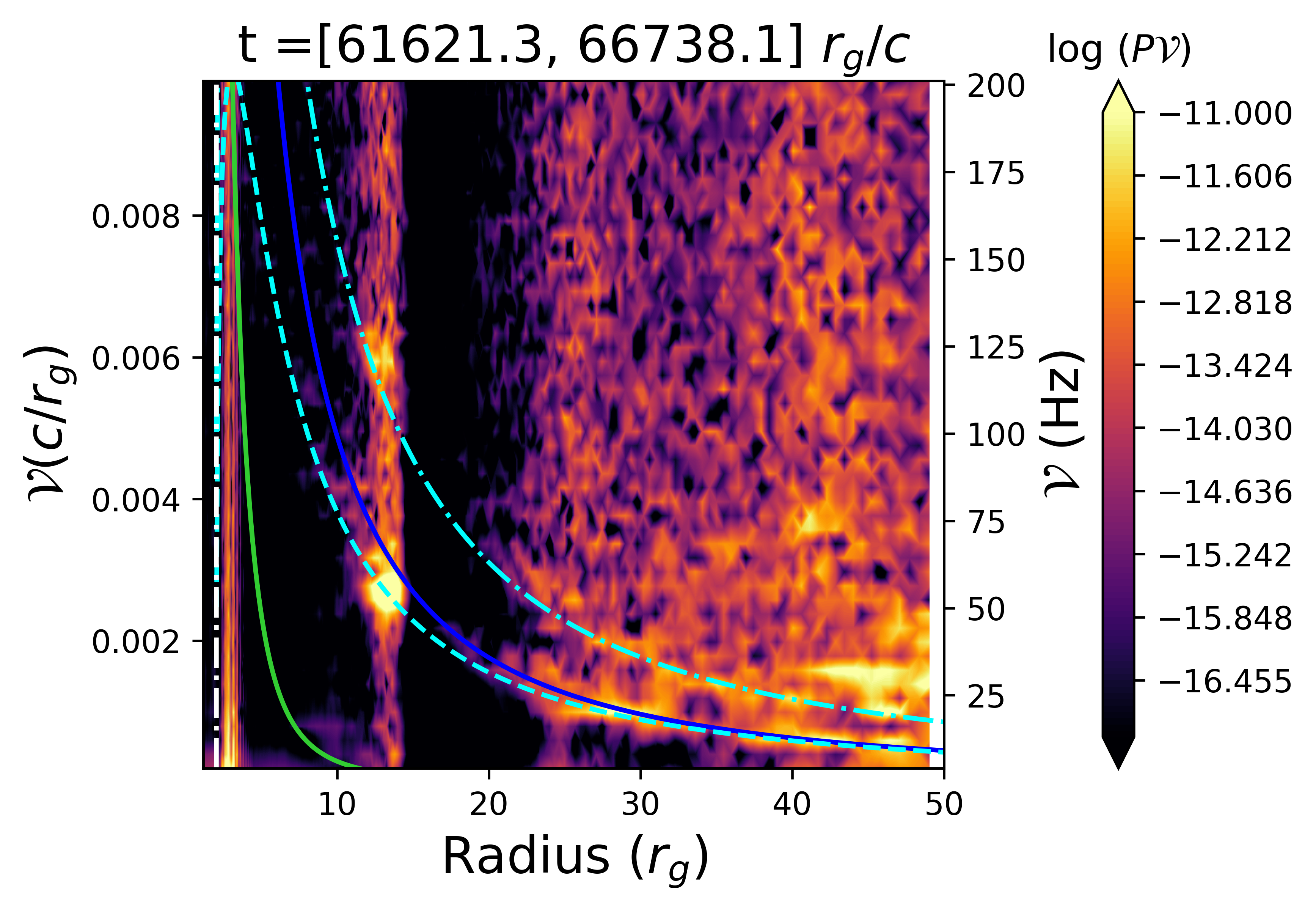}
    \includegraphics[clip,trim=0.0cm 0.0cm 0.0cm 0.0cm,width=7.0cm]{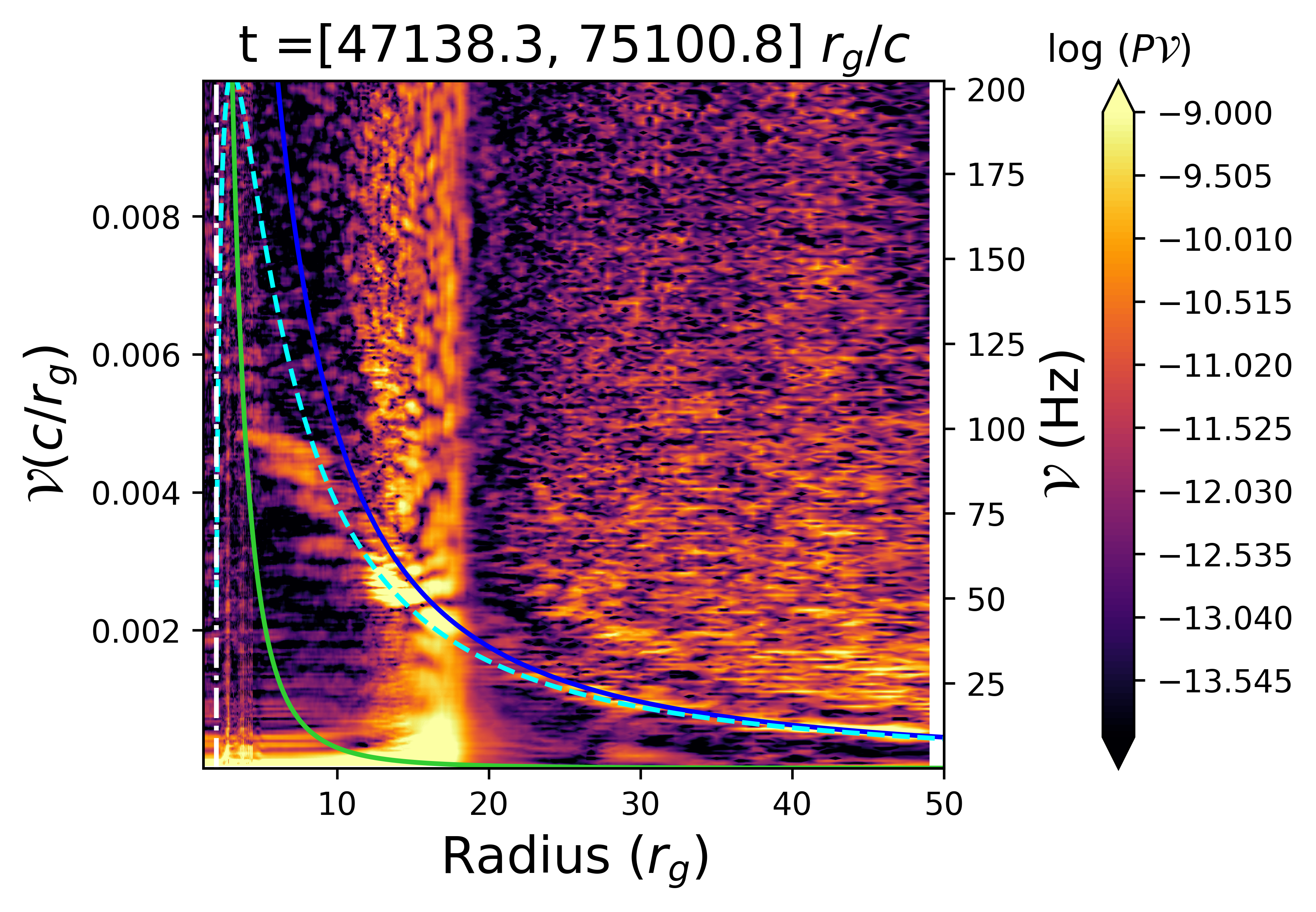}
    \includegraphics[clip,trim=0.0cm 0.0cm 0.0cm 0.0cm,width=7.0cm]{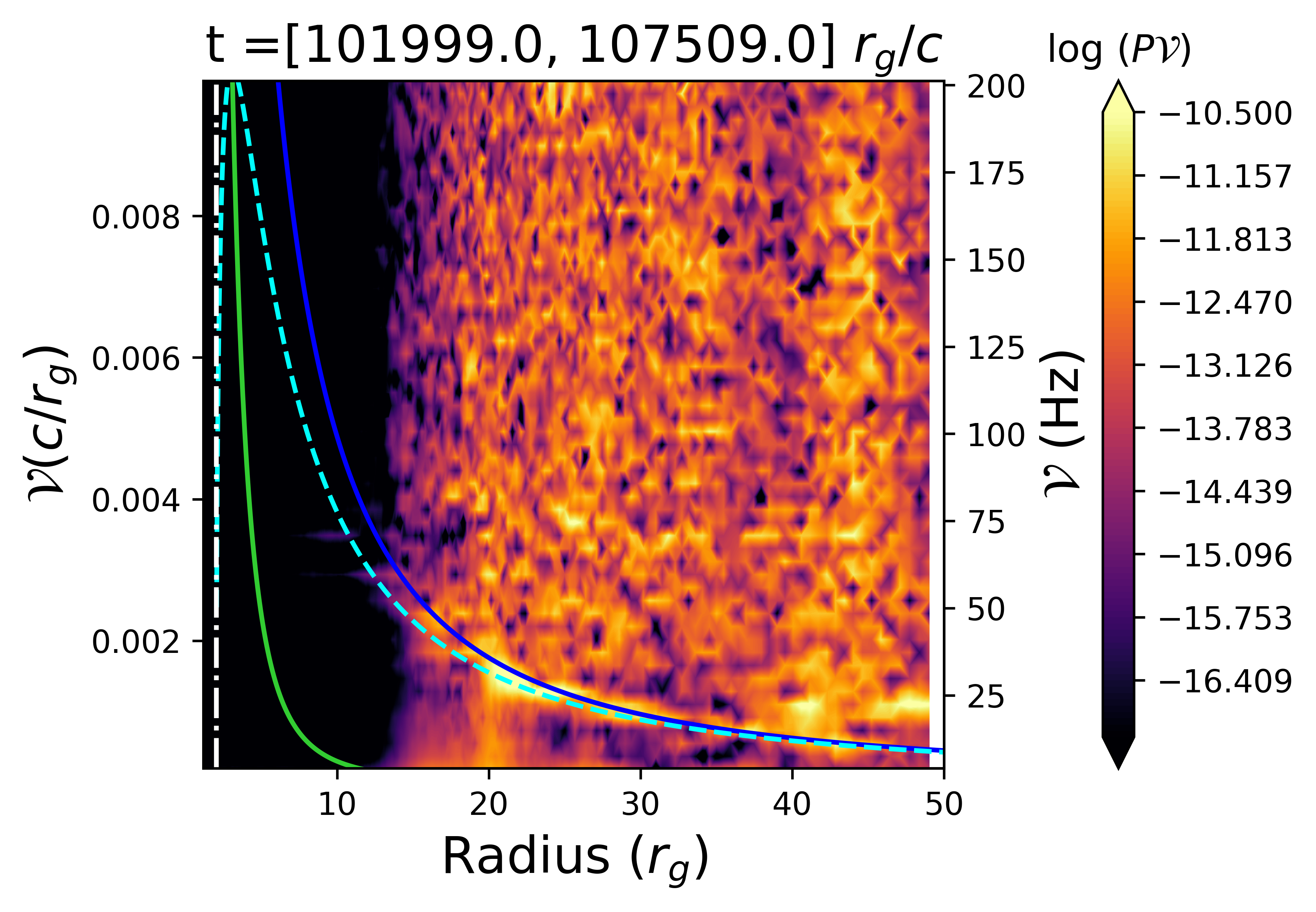}
    \caption{PSDs of the radial mass flux $\dot{M}$ for the inner $50r_g$ of the disk. In each figure the dashed cyan, solid blue and green curves correspond to the radial epicyclic, Keplerian and Lense-Thirring precession frequencies respectively. The colour bar shows log of the spectral power ($P \times \nu$). The dash-dot cyan curve in the top right panel corresponds to 2 times the radial epicyclic frequency. The vertical white dash-dot curve in each plot shows the location of the ISCO ($=2.044r_g$). The frequency $\nu$ in Hz (right $y$-axis in each plot) is computed for a 10 $M_\odot$ BH. The time period over which each PSD is computed is given above each plot. \textbf{Top:} PSD computed during the largest tearing cycle E7. The time period over which the PSD is computed captures the period during which the inner sub-disk begins to undergo Bardeen-Petterson alignment, in which the inner $5r_g$ of the inner sub-disk begins to align with the BH spin axis. Extended regions of enhanced power (yellow) are found along 1 and 2 times the radial epicyclic frequency curves (dashed and dash-dot cyan curves respectively). Along these extended high power regions lies two localised regions of high power, each signifying the presence of a HFQPO, are found at the position of the tearing radius during the given time period ($\sim13r_g$). The HFQPOs have centroid frequencies of $\sim55$ and $110$Hz, which roughly coincide with 1 and 2 times the radial epicyclic frequency at $13r_g$. \textbf{Middle:} PSD computed over a larger time window during the tearing cycle E7. While the 55 Hz HFQPO signature remains present the weaker 110 Hz feature is smeared out and the extended regions of high power become difficult to identify. \textbf{Bottom:} PSD computed during a time period in which no disk tearing occurs. There is a notable suppression of the spectral power within the inner $\sim13r_g$ of the disk.}
    \label{fig:Mdot}
\end{figure}

To identify oscillatory signals in the simulation, we use the radial mass flux $\dot{M}$ given in equation \ref{eq:Mdot}. In order to probe the structure of the variability in the simulation, we compute the power spectral density (PSD) of the radial mass flux. For each spherical shell in the numerical grid, the PSDs are computed similarly to \cite{Hogg_Reynolds2016} using $P(\nu) = \mid \tilde{\dot{M}}(\nu) \mid ^{2}$, where $\tilde{\dot{M}}(\nu)$ is the Fourier transform of the time-series (in a given radial shell) of the radial mass flux $\dot{M}$. 
In practise, we apply the numpy fft function which implements a discrete Fourier transform \citep{CooleyTukey} and the reported PSDs are proportional to the variance of the signal. 
The time-series have a cadence of $50 r_g/c$ resulting in a Nyquist frequency of $203\, \rm Hz$ (for a $10\, M_{\odot}$ black hole). In Figure \ref{fig:Mdot} we show PSDs across spherical shells for the inner $50r_g$ of the disk, for time periods in which the disk does not tear (bottom panel) and at times during the longest tearing cycle E7 (top two panels). We also overplot curves showing the characteristic frequencies \citep[e.g.][]{Nowak_Lehr1998} on top of the PSDs in Figure \ref{fig:Mdot}; the Keplerian frequency (solid blue curve) given by  
\begin{equation}
\nu_{\Omega} = \frac{1}{2 \pi} \left( r^{3/2} + a\right)^{-1} \text{,}
\end{equation}
the radial epicyclic frequency (dashed cyan curve) given by
\begin{equation}
\nu_{r}^{2} = \nu_{\Omega}^{2} \left(1 - \frac{6}{r} + \frac{8a}{r^{3/2}} - \frac{3a^{2}}{r^{2}}\right) \text{ ,}
\end{equation}
and the Lense-Thirring precession frequency (solid green curve)
\begin{equation}
\nu_{LT} = \frac{1}{2 \pi}\frac{2a}{r^{3}} \text{ ,}
\end{equation}
where $a = 0.9375$ is the black hole spin and $r$ is the radius and units are set such that $c = G = 1$ \footnote{We note that the characteristic frequencies listed here are for aligned accretion flows. Approximate expressions for inclined orbits have been presented by \cite{Sibgatullin2001}. We have used the available relations to check that for the typical tearing radii (e.g. $\gtrsim 10r_g$), inclination leads to a negligible shift of the Keplerian and radial epicyclic frequencies (in fact away from each other).}.  

All PSDs show signatures of enhanced variability throughout most of the disk near radial epicyclic or nearby Keplerian frequencies (shown as cyan dashed- and blue solid lines which are nearby at large radii).
In the absence of disk tearing, we find a depression in the spectral power within the inner $\sim18r_g$ of the disk, with regions of highest power coinciding with larger radii ($r>18$) and low frequency.  The low central power in absence of tearing is consistent with the low densities in the inner regions during these time intervals (c.f. Figure \ref{fig:spacetime}).  
During the disk tearing cycle E7, we find an enhancement in the spectral power across all frequencies at radii close to the position of the tearing radius ($\approx13r_g-14.5 r_g$ during time window in which the PSD is computed). Importantly, the epicyclic signal at the tearing radius is boosted most strongly, giving rise to a 55 Hz signal if scaled to a $10 M_\odot$ BH (bright yellow feature at $r\approx13r_g$ and 55 Hz). This feature, which we identify as a HFQPO, is clearly seen in both the top two panels of Figure \ref{fig:Mdot}, and is notably absent in the PSD in which the disk does not tear (bottom panel of Figure \ref{fig:Mdot}).  

\begin{figure}
    \centering
    \includegraphics[clip,trim=0.0cm 0.0cm 0.0cm 0.0cm,width=0.46\textwidth]{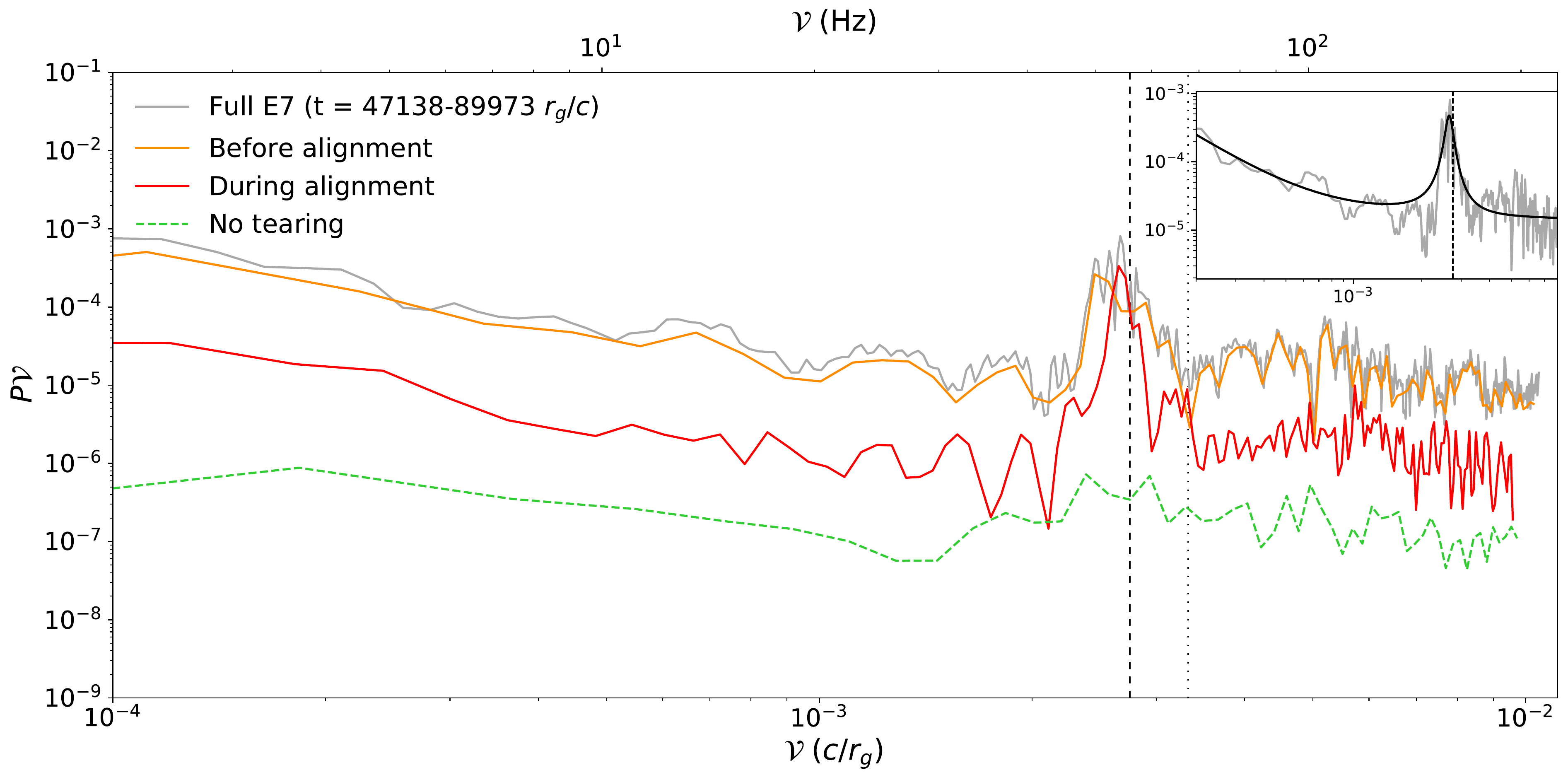}
    \caption{Power spectra computed for different time intervals during tearing cycle E7 (except the dashed green curve). All power spectra shown are averaged over three discrete radii, 13, 14 and 15 $r_g$. These radii correspond to locations at or near the position of the tearing radius during the time period in which the HFQPOs are present in tearing cycle E7. The lower $x$-axis shows the frequency in units of $c/r_g$ while the upper $x$-axis shows frequency in Hz computed for a 10 $M_\odot$ BH. The spectral power $P \times \nu$ is computed with frequency $\nu$ in units of $c/r_g$. The grey curve shows the power spectrum computed over the entire cycle E7. The solid red and orange curves show the power spectra computed prior to any alignment of the inner sub-disk (orange, $t= [47138,56200] r_g/c$) and during alignment (red, $t= [56200,72841] r_g/c$) in cycle E7. The dashed green line shows the power spectrum computed for a period in which no disk tearing occurs ($t = [101999,107509] r_g/c$). The vertical black dashed line shows radial epicyclic frequency at 13$r_g$, which is found to roughly coincide with the centroid frequency of the 55 Hz QPO peak. The vertical black dotted line shows the Keplerian frequency at 13$r_g$. The HFQPO feature seems to be strongest during the global alignment phase. The inset shows the same power spectra computed over the entire tearing cycle E7 (grey curve) alongside the corresponding fit (black curve) obtained with a Lorentzian and two power-laws. The fit gives QPO quality factor $Q = 11.3$.}
    \label{fig:multi-powerspec_BP}
\end{figure}

\begin{figure}
    \centering
    \includegraphics[clip,trim=0.0cm 1.0cm 1.0cm 1.0cm,width=8.3cm]{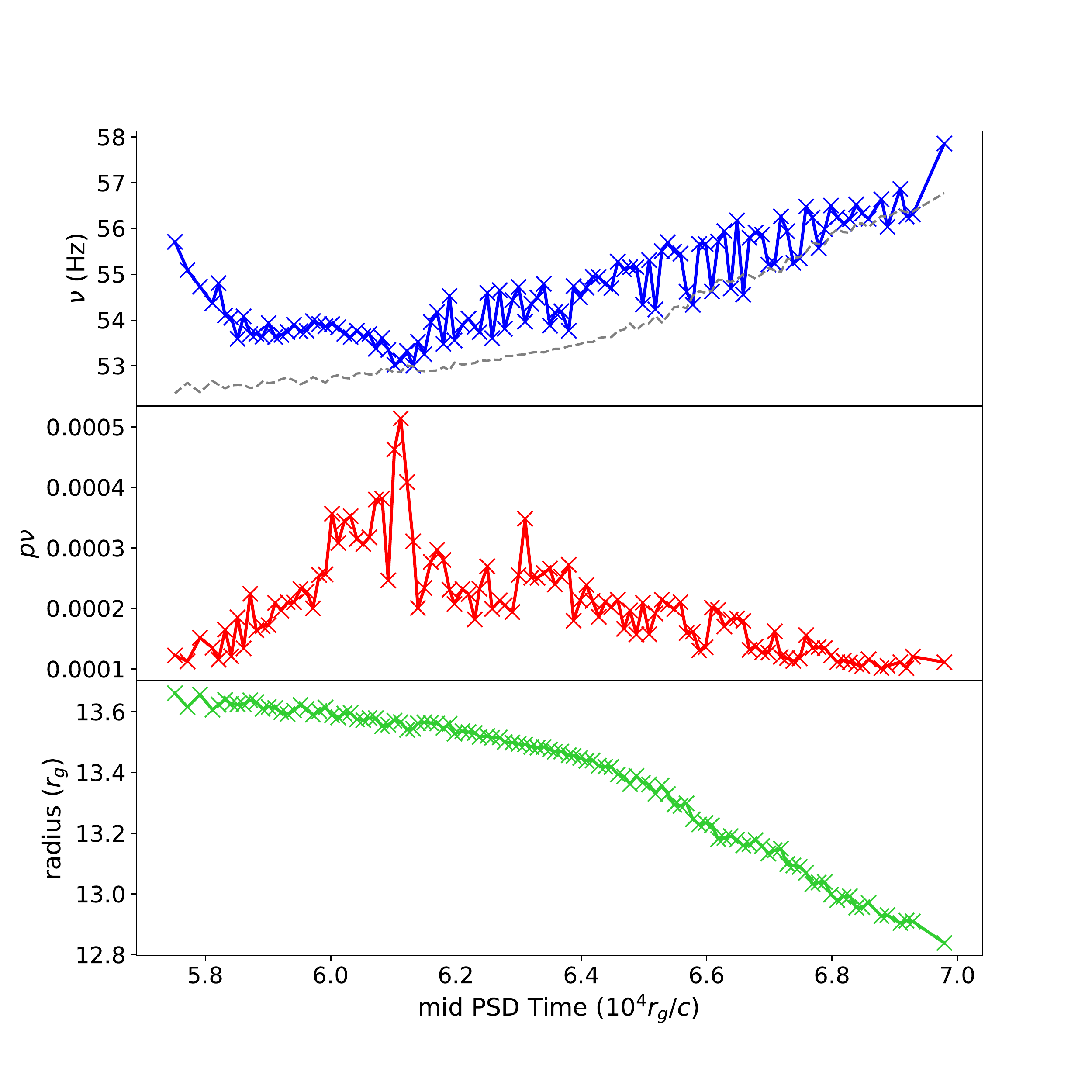}
    \caption{Evolution of the 55 Hz (fundamental) HFQPO feature as a function of time. \textbf{Top:} HFQPO centroid frequency as a function of time is shown by the blue curve. The dashed grey curve shows the radial epicyclic frequency at the radial location of the HFQPO peak in the 2D PSDs; note that the corresponding Keplerian frequency would be outside the plotted range of frequencies. \textbf{Middle:} HFQPO spectral power ($p \times \nu$, with frequency $\nu$ in units of $c/r_g$). \textbf{Bottom:} radius at which the HFQPO is triggered. These parameters are obtained by fitting a 2D Gaussian to the 55 Hz HFQPO feature in 2D PSDs of the radial mass flux $\dot{M}$ (e.g. such as those shown in Figure \ref{fig:Mdot}). The fits are performed on a series of sequential PSDs in order to find how the fit parameters evolve over time. The mid PSD time on the $x$-axis corresponds to the centre point of the time window used to compute each 2D PSD. The curves shown roughly span the time period over which the HFQPO is stable.}
    \label{fig:freq_tracking}
\end{figure}

Figure \ref{fig:multi-powerspec_BP} shows power spectra averaged over three discrete radii, 13, 14 and 15 $r_g$, for cycle E7. These radii correspond to locations at or near the position of the tearing radius during the time period in which the HFQPOs are present in tearing cycle E7. The HFQPO clearly stands out as a distinct peak with a centroid frequency of $\sim55$Hz and a quality factor of $Q = 11.3$. The centroid frequency is well described by the radial epicyclic frequency at $13 r_g$ (dashed vertical line, i.e. roughly the current position of the tearing radius). In Figure \ref{fig:freq_tracking} we show the evolution of the frequency, spectral power $P \times \nu$ and radius at which the 55 Hz HFQPO is triggered as a function of time. We track the evolution by fitting a 2D Gaussian to the 55 Hz HFQPO feature in the 2D PSDs of the radial mass flux (such as those shown in Figure \ref{fig:Mdot}). The PSDs used for the fits are computed using a fixed time window of size $5\,117 r_g/c$ (same as the top panel of Figure \ref{fig:Mdot}), and are computed every $100 r_g/c$ across a time period roughly corresponding to the time during which the 55 Hz HFQPO is a stable feature in the PSDs. The centroid frequency of the 55 Hz QPO (top panel of Figure \ref{fig:freq_tracking}) varies by only a few percent, increasing over its lifetime, and is found to have an average frequency of $54.7$ Hz over this time. 

The bottom panel of Figure \ref{fig:freq_tracking} shows that the radial position at which the HFQPO is triggered moves inwards towards the BH, by $\sim 1 r_g$, correlated with the small increase in the centroid frequency of the HFQPO over its lifetime. To relate the HFQPO to disk tearing, we inspect the tilt angle for rapid changes in the radial location of the tearing radius (see e.g. Figure \ref{fig:tilt_v_rad}). The disk initially tears at $\sim 20 r_g$ at the start of the tearing cycle E7. The tearing radius then moves inwards until it settles between $\sim13-15.6 r_g$ (shown by the vertical grey band in \ref{fig:tilt_v_rad}) for the entire period in which the HFQPO signatures are present. The small variation in tearing radius is consistent with the small variation found for the radius at which the HFQPO is triggered, thus highlighting the dependence of the evolution of the HFQPOs centroid frequency, and location at which it is triggered, on the evolution of the disk tearing radius. 

We find that the spectral power $P \times \nu$ (middle panel Figure \ref{fig:freq_tracking}) initially increases over time before reaching a maximum, which may indicate that the triggering mechanism takes some time to setup. The spectral power of the HFQPO then begins a slow decay back to a power similar to what the HFQPO had when it was first stable. 

Comparing the $55$ Hz HFQPO signal in Figure \ref{fig:multi-powerspec_BP} to the power spectrum computed for times during which the disk does not tear (green curve in Figure \ref{fig:multi-powerspec_BP}), we find that spectral power is strongly enhanced during disk tearing. The QPO signature is present during disk tearing regardless of whether the inner sub-disk is partly aligned with the black hole or not (orange and red curves), however, as also illustrated in figure \ref{fig:tilt_v_rad} the tear moves in rapidly once partial alignment has reached the tearing radius $r\approx 13 r_g$. This marks the end of the HFQPO signatures in the power spectra and PSDs.  

Interestingly, we also see evidence for a, albeit weaker, $\times 2$ harmonic signal throughout the disk. During disk tearing, this harmonic is boosted to result in a weaker HFQPO at $\sim110$ Hz (see localised yellow region at 110 Hz along the dash-dot cyan curve in in top panel of Figure \ref{fig:Mdot}). The 110 Hz HFQPO has weaker spectral power compared to the 55 Hz feature, as can be seen in the top panel of Figure \ref{fig:Mdot}. Thus the 110 Hz QPO becomes more difficult to identify when the PSDs are computed over large time windows (e.g. middle panel of Figure \ref{fig:Mdot}) and when averaging the power spectra over multiple radii (e.g. Figure \ref{fig:multi-powerspec_BP}). Similarly to the $55$ Hz peak, the $110$ harmonic HFQPO is only found to vary by a few Hz during its lifetime. The presence of simultaneous HFQPO peaks with a 1:2 ratio found in this work is consistent with the observed ratios of HFQPOs \citep[e.g. GRS 1915+105 ][]{Belloni_Altamirano2013}, however, we have not observed the more common 2:3 resonance in the PSDs of the radial mass flux. 

\begin{figure}
    \centering
    \includegraphics[clip,trim=0.0cm 0.0cm 0.0cm 0.0cm,width=0.45\textwidth]{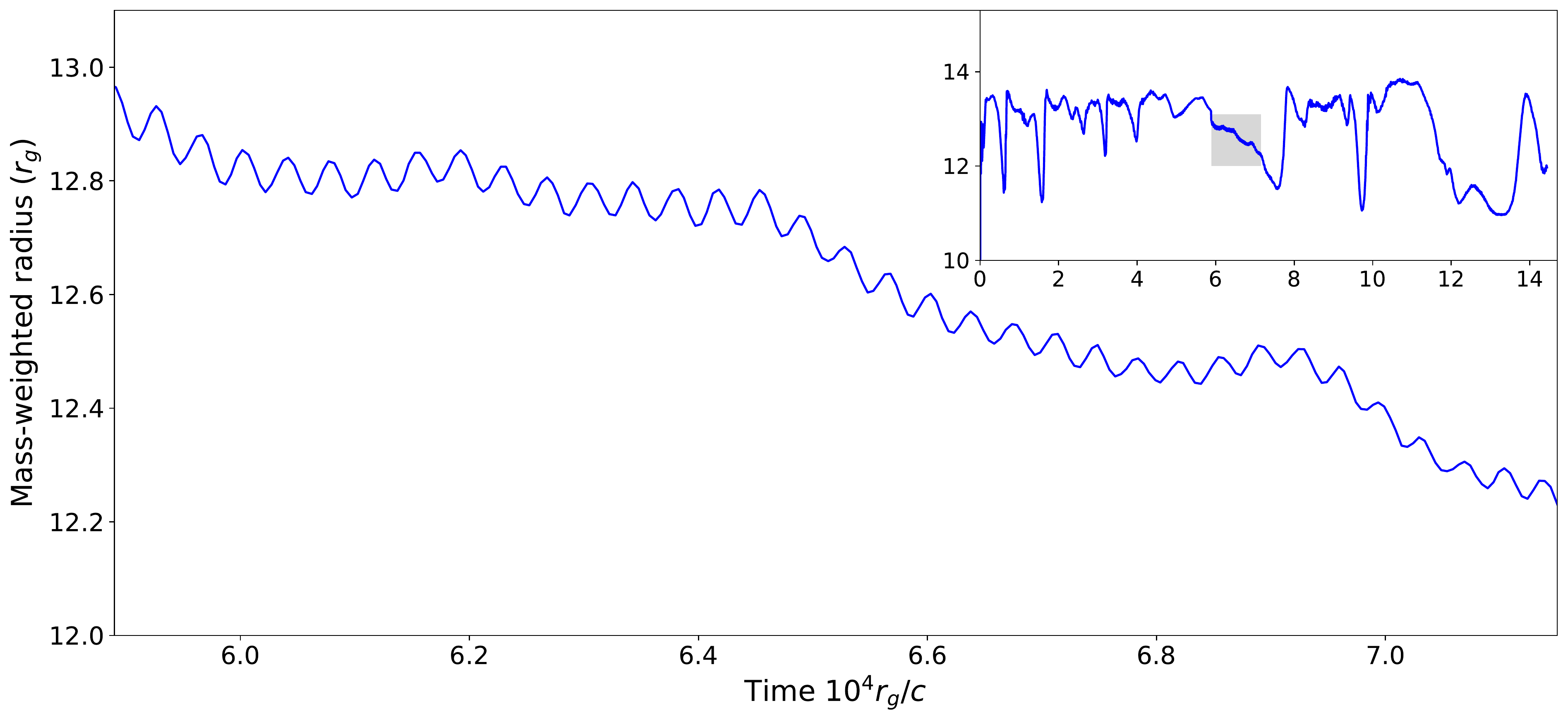}
    \includegraphics[clip,trim=0.0cm 0.0cm 0.0cm 0.0cm,width=0.45\textwidth]{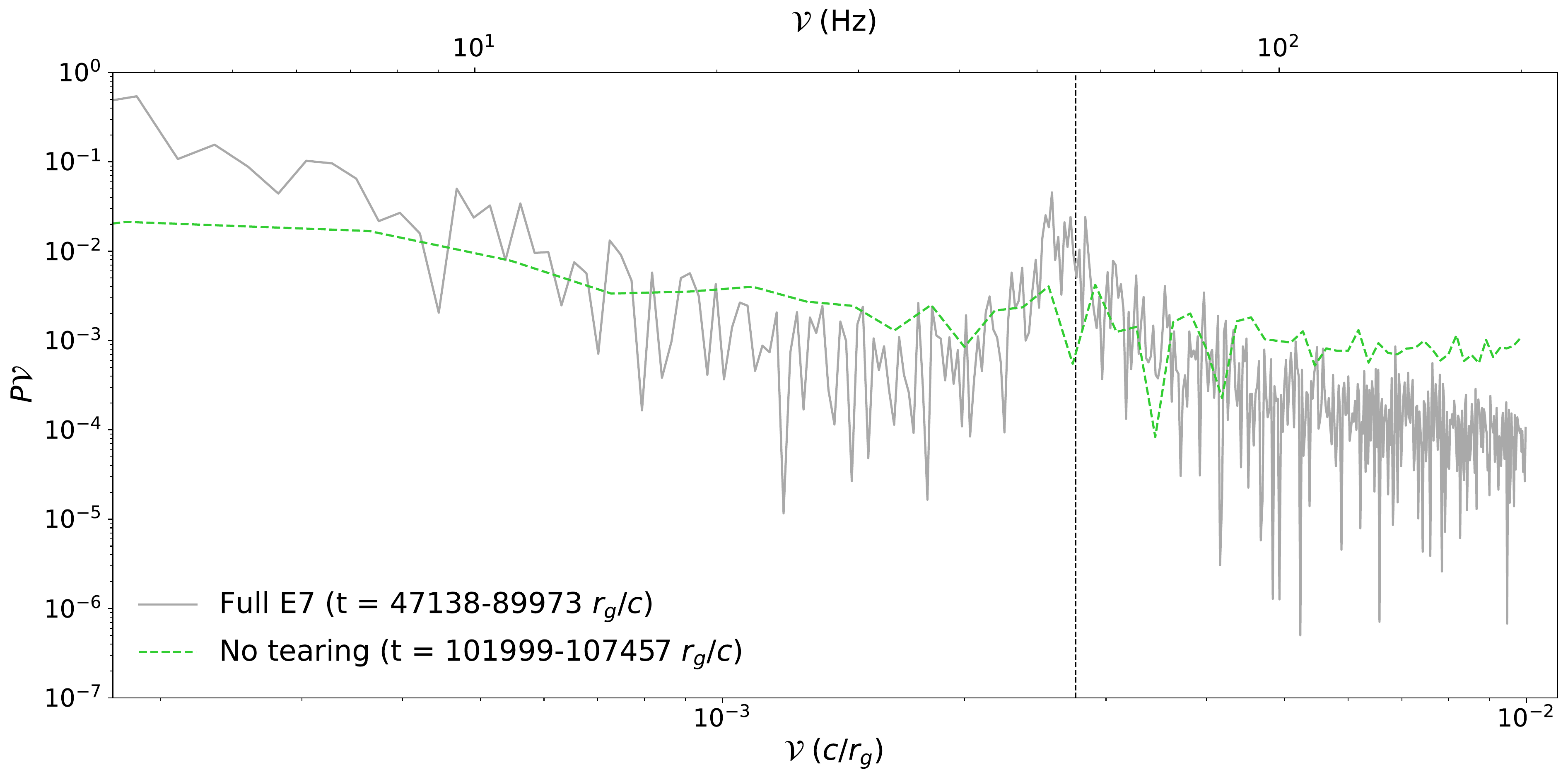}
    \caption{Physical oscillations of the inner sub-disk. \textbf{Top:} Mass-weighted radius as a function of time. The inset shows the full mass-weighted radius for the full simulation time, the main plot shows a zoom in of the region shown by the grey rectangle in the inset plot.  \textbf{Bottom:} Power spectra of the mass-weighted radius computed over the full tearing cycle E7 (grey curve) and a period of time in which the disk does not tear (dashed green curve), as in Figure \ref{fig:multi-powerspec_BP}. The vertical dashed black line shows the radial epicyclic frequency at $13 r_g$ ($55\rm Hz$).}
    \label{fig:Barycen_powerspec}
\end{figure}

The variations in the radial mass flux are indicative of a radial oscillation or ``breathing mode'' of the inner disk. This feature can be seen in the animation of the vertically integrated disk density provided on the following \href{https://www.youtube.com/playlist?list=PLDO1oeU33GwlE5_hWjOq0FgDx6I_z856Y}{YouTube channel}. The inner sub-disk has formed a pronounced dense ring extending out to the tearing radius where rapid oscillations with a period corresponding to the $55$ Hz HFQPO can be identified. The ring shows no discernible eccentricity. 
To quantify this motion further, we compute the mass-weighted radius as 
\begin{align}
    r_{\rm B} = \frac{\int_{r_1}^{r_2} dr \int d\Omega \, r\, \rho(r,\theta,\phi)}
    {\int_{r_1}^{r_2} dr \int d\Omega \rho(r,\theta,\phi) }
\end{align}
where the radial integration is performed between $r_1=10r_g$ and $r_2=15r_g$ to capture the ring.  Figure \ref{fig:Barycen_powerspec} shows the time evolution of $r_{\rm B}$ and its associated PSD.  The $55\rm Hz$ oscillations seen before in the radial mass flux are clearly present in the motion of the barycenter, indicating that the ring performs a global $m=0$ breathing oscillation.  No prominent frequency is observed in the control volume in the absence of tearing (see green curves for comparison).  

\subsection{Detection of a LFQPO of the inner sub-disk} \label{sec:LFQPO_detection}

\begin{figure}
    \centering
    \includegraphics[clip,trim=0.0cm 0.5cm 1.0cm 1.8cm,width=0.48\textwidth]{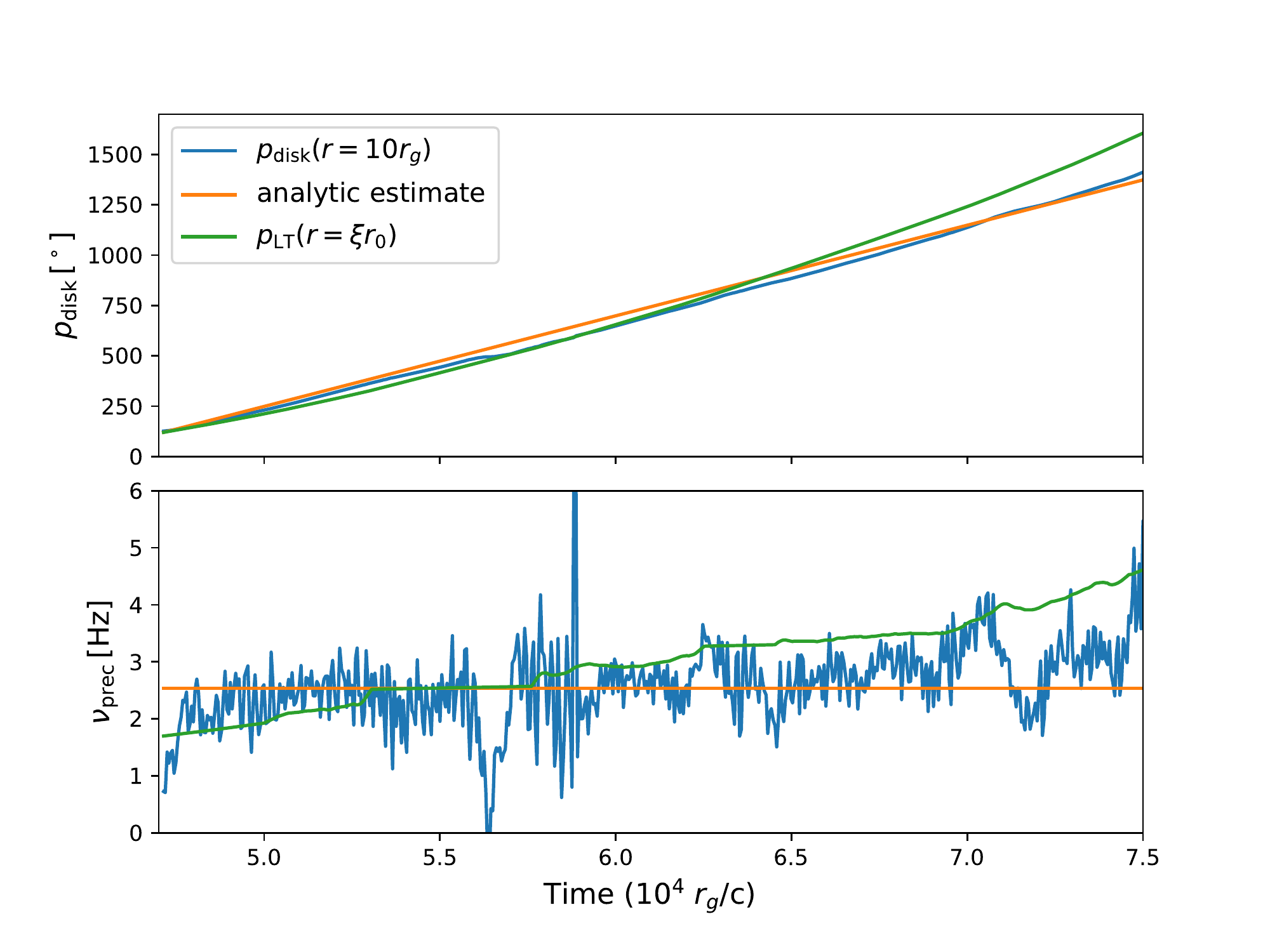}
    \caption{Precession of the inner sub-disk for cycle E7.  The actual values are extracted at $r=10\, r_g$ shown as blue lines.  We also show the analytic estimate base on the integrated misaligned angular momentum (orange) and the nodal precession for a ring with radius $\xi r_0$ where we set $\xi=0.75$ and $r_0$ denotes the instantaneous tearing radius (green). The top panel shows the cumulative precession angle and the bottom panel the instantaneous frequency in Hz for a $10 M_{\odot}$ black hole.}
    \label{fig:precession}
\end{figure}

Once a tear has developed, the inner and outer disk precess independently with a frequency given by the local Lense-Thirring (LT) torque and angular momentum of the sub-disk  \citep[e.g.][]{Liska_Tchekhovskoy_Ingram2019_BP2,Liska_Hesp_Tchekhovskoy2021_disc_tearing_BP}.  This leads to a rapid precession of the inner disk as seen in the precession angle $p_{\rm disk}$ shown in Figure \ref{fig:spacetime}. Since the inner disk is warped, a phase lag occurs where the inner parts ($\sim 5 r_g$) lead before the outer parts. However, the precession frequency is essentially constant throughout the entire sub-disk. Thus we characterise the precession as \textit{warped solid body}.  
There are several further important observations to be made: a) the precession increases in frequency as the tear moves inwards, b) the density of the disk is highly peaked towards its edge suggesting that the precession frequency can be approximated from test particle motion at the tearing radius.  
We test this hypothesis in Figure \ref{fig:precession}, where we compare the phase $p_{\rm disk}$ of the sub-disk with the analytic prediction from the nodal precession frequency $\nu_{\rm LT}$ \citep[see e.g.][]{LewinvanderKlis2010} $p_{\rm LT}(r)=\int 2\pi \nu_{\rm LT}(r) dt$.  

The actual precession angle of the disk has been extracted at $r=10\, r_g$, however, due to the solid-body nature of the sub-disk, the radius does not matter much.  
We overlay two curves: first, we show an analytic estimate of the precession frequency based on the integrated misaligned angular momentum divided by the integrated Lense-Thirring torque as in \cite{Liska_Tchekhovskoy_Ingram2019_BP2}. This results in a period of $8000 r_g/c$ which recovers the  average period with very good accuracy. Second, we show the nodal precession with ``effective radius'' $\xi r_0$ where $r_0$ is the instantaneous tearing radius and the parameter $\xi<1$ is a coefficient that accounts for the moment of inertia of the inner disk.  If the ring were infinitely thin and located at $r_0$, we would have $\xi=1$.  We find that the actual precession behaviour is quite well captured by $\xi=0.75$.  
As the time progresses, the frequency increases from $\sim2\rm Hz$ to $\sim 4\rm Hz$ which is quite well captured by the model that takes the shrinking of the tearing radius into account.

\section{Discussion} \label{sec:discussion}

In our simulation, disk tearing is associated with the formation of rings with enhanced density and we observe that a tear always occurs behind such a ring.  For the large tearing cycles E7 and E10, the ring lasts for the majority of the tear (around $\simeq20,000 r_g/c$) and the tearing radius remains nearly constant. During this time, the inner disk performs several full precession cycles and many radial epicyclic oscillations. This results in well defined HFQPO and LFQPO signals which we extract from the PSDs of radial mass flux and disk precession angle respectively. We will now discuss our findings in the context of previous work and in relation to X-ray timing observations.  

\subsection{Comparisons to previous work}
A number of numerical studies have previously been performed to look for evidence of QPOs in simulations, with a few showing some evidence for the generation of QPOs. For example \cite{Kato2004} find a promising signature of twin HFQPOs in their pseudo-Newtonian MHD simulations thought to be produced by the resonance between the epicyclic and Keplerian frequencies. \cite{Schnittman_Krolik_Hawley2006} found excess power by ray tracing a three dimensional global GRMHD simulation of an accreting Schwarzschild black hole indicative of a HFQPO pair with a 2:3 frequency ratio. However, since this excess power is orders of magnitude weaker than in observed HFQPOs, this work highlights the need for further simulations. 

Evidence for epicyclic modes in tilted disks was previously found by \cite{Fragile_Blaes_Omer2008} who analysed a GRMHD simulation of a ($15^{\circ}$) tilted, geometrically thick ($\alpha<H/r$) disk around a Kerr black hole. In their study, it is argued that epicyclic modes are driven by pressure gradients attributed to the warping of the disk. \cite{Henisey_Blaes_Omer2009} analyse this simulation further and find enhanced (radial) bands of power at frequencies below the orbital frequency, which they interpret as inertial (or acoustic) modes. It is argued that these modes could in principle explain the frequency range of HFQPOs in BHXRBs, but these authors do not find a clear HFQPO signature. The most important difference to our study is that the disk of \cite{Fragile_Blaes_Omer2008, Henisey_Blaes_Omer2009} is much thicker and does not tear.

The radially extended regions of high power along the radial epicyclic curve shown in Figure \ref{fig:Mdot} have similarly been found in the PSDs (e.g. of the fluid three velocity) generated from the viscous, radiative hydrodynamic 2D simulation of aligned thin disks around a Schwarzschild black hole analysed in \cite{Mishra_Kluzniak_Fragile2019,Mishra_Kluzniak_Fragile2020}. In \cite{Mishra_Kluzniak_Fragile2019} they find radially extended regions of high power along the radial epicyclic curve. \cite{Mishra_Kluzniak_Fragile2020} also report local enhancements of the variability manifested as ``stripes'' in the PSDs, one of which is identified as a trapped g-mode (their Figures 1\&4). These features are promising, however it remains to be seen which of the numerous modes are robust against damping by the magnetorotational-driven turbulence.

A model that bears some resemblance with our simulation was proposed by  \cite{Fragile_Mathews_Wilson2001} who postulate that the Bardeen-Petterson effect in thin tilted disks may result in the generation of QPOs via material passing through the transition region (Bardeen-Petterson transition radius) between the resulting aligned inner and misaligned outer sub-disks. If the midplanes of the inner and outer sub-disks differ by moderately large angles ($\theta \ge 30^{\circ}$) then the sub-disks are almost disconnected. It is speculated that material can then be accreted from the outer sub-disk to the aligned inner sub-disk via a tenuous stream of matter or in the form of clumps of gas, both of which may lead to shock heating when this accreted material hits the outer edge of the inner-sub disk. In this picture, the shock heated gas that then orbits the black hole at a radius just inside the Bardeen-Petterson transition radius, is responsible for generating QPOs with frequencies close to the Keplerian orbital frequency. 

While some of the features just described are borne out of our simulation, we find that the dynamical evolution unfolds differently: rather than forming an outward-moving alignment radius, our disk undergoes cyclic tearing events.  During these cycles, we find the inner disk does indeed align with the black hole, however, the alignment in itself cannot be responsible for the QPO generation as the  the QPOs are triggered before alignment occurs. This is seen by the presence of the 55 Hz HFQPO peak in the orange curve in Figure \ref{fig:multi-powerspec_BP}, in which the PSD is computed during a time period prior to Bardeen-Petterson alignment.
Our simulation, instead, shows that the signal (with radial epicyclic-, not keplerian- frequencies) is associated with the tearing radius and likely caused by differential precession between the two disks.

\subsection{Towards a unified QPO mechanism}
Disk tearing in tilted accretion flows has the potential to unify QPO production in BHXRBs, provided that precession driven LFQPOs can be produced (via disk precession e.g. \citealp{StellaVietri1998,IngramEtAl2009,Ingram_van_der_Klis_2015,Miller_Homan2005,Fragile_Anninos2005,Veledina_Poutanen_Ingram2013} and/or jet precession driven by the precession of the innermost regions of the accretion flow \citep{Liska_Hesp_Tchekhovskoy2018} e.g. type-B LFQPOs \citealp{Liska_Hesp_Tchekhovskoy2018,Kylafis_Reig_2019}, see also \citealp{Stevens_Uttley2016,Kalamkar_Casella_Uttley2016}). In the popular relativistic precession model for LFQPOs proposed by \cite{IngramEtAl2009}, the generation of a type-C LFQPO is attributed to the solid body precession of the hot, geometrically thick inner flow of a truncated accretion disk. In this picture the frequency of the LFQPO is given by the Lense-Thirring precession frequency and the frequency evolution governed by the evolution of the truncation radius-- as the truncation radius moves inwards during an outburst, the precession increases causing the LFQPO to increase. In the simulation we present here the onset of disk tearing leads to the self-consistent generation of a precessing inner sub-disk that undergoes solid body precession and an analysis of the precession angle reveals that the inner sub-disk undergoes nearly 5 precession periods with a period of $\approx 8\, 000r_g/c$. This  frequency indicates that the inner sub-disk could potentially produce a precession driven LFQPO with a frequency of $2.5$Hz (for a $10 M_\odot$ black hole), consistent with the range of LFQPO frequencies predicted by e.g. the Lense-Thirring precession driven model of \cite{IngramEtAl2009}. Thus a LFQPO could be triggered during a time window that overlaps with that in which the HFQPOs are observed (the HFQPO signatures are seen during the time window $\sim56\,000 - 74\,000$ $r_g/c$).  As discussed further in Appendix \ref{sec:measuring-spin}, such a simultaneous detection would allow one to put constraints on the black hole spin assuming the black hole mass is known through other means.  

\subsection{Observational considerations}
\label{sec:observational_considerations}
We have demonstrated that disk tearing followed by LT-precession of the inner disk self-consistently explains a single LFQPO peak in addition to a HFQPO peak and its harmonic. While this is very promising, XRBs typically display a rich phenomenology of Type-A, Type-B and Type-C LFQPOs. The absence of Type-B QPOs is expected, since Type-B QPOs might be associated with a jet (e.g. \citealp{Remillard_McClintock2006, Ingram_Motta2019, Marcel_2020SADJET}), which is absent in our simulation. GRMHD Simulations have already demonstrated that jets can precess and lead to quasi-periodic emission \citep{Liska_Hesp_Tchekhovskoy2018}. In addition, they have demonstrated that during tearing events the jet can get deflected when it crosses the midplane of the outer disk, which can lead to interesting phenomenology not captured in this work \citep{Liska_Hesp_Tchekhovskoy2021_disc_tearing_BP}. 

One potential problem in our model is that the longest tearing cycles have life and recurrence times of $\sim 1-3s$, which is extremely short compared to the time during which QPOs are detected. In addition, the disk in our simulation tears over a range of radii ($10-20 r_g$), which leads to a broad spectrum of periods. At first sight this seems to be problematic for our model. However, recent observations have shown that LFQPOs decohere after a few cycles \citep{van_den_Eijnden_Ingram_Uttley2016}, which would be expected if QPO emission is in fact made up out of many stochastically distributed tearing events. In addition, tearing events occurring at much smaller radii than the fundamental radius of $r \sim 13 r_g$ live very short and rarely complete more than one precession period. For example, tearing events E7 and E10 take up half of the simulation time, while all of the other tearing events combined only take up $10\%$ of the simulation time. Of these other tearing events only E2, E3 and E9 complete more than one precession period. Longer duration GRMHD simulations, which establish mass inflow equilibrium to a much larger radius, will be needed to verify that disk tearing will not lead to more than a modest broadening of the observed QPO peaks.

\begin{figure}
    \centering
    \includegraphics[clip,trim=0.0cm 0.0cm 0.0cm 0.0cm,width=0.45\textwidth]{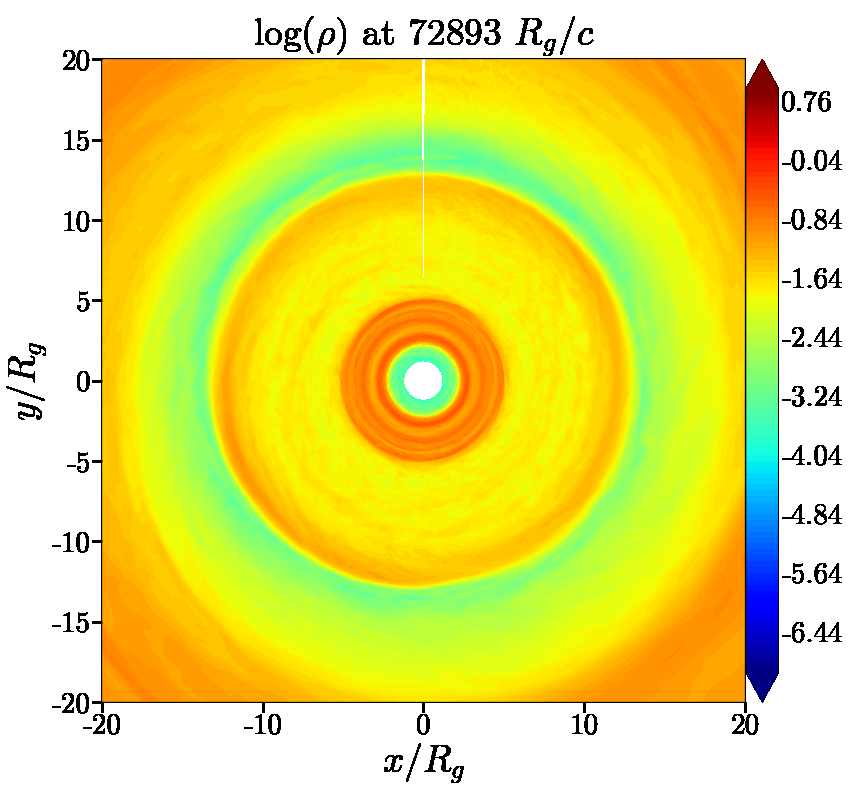}
    \caption{Top down view of the disk showing $\theta-$integrated integrated density. The inner disk exhibits concentric high density rings between the ISCO and Bardeen-Petterson alignment radius $r_{BP} \sim 5 r_g$. We speculate in section ~\ref{sec:observational_considerations} that such rings might be associated with disco-seismic oscillations, which are responsible for HFQPO pairs above the maximum radial epicyclic frequency (e.g. \citealp{Kato2004}).}
    \label{fig:rings}
\end{figure}

One particularly challenging aspect is the detection of a HFQPO peak at 450 Hz in GRO J1655-40, which, most likely, lies above the maximum radial epicyclic frequency \citep{Strohmayer2001a}. An intriguing possibility is that different physical mechanisms can be responsible for different HFQPOs. Namely, our work does not exclude the possiblity that some HFQPOs are caused by trapped discoseismic p- and g- modes (e.g. \citealt{Kato2004}) or other physical mechanisms (e.g. \citealt{Kluzniak_Abramowicz2002}). In fact, figure \ref{fig:rings} shows the formation of concentric, high density rings in the BP-aligned portion of the disk within our simulations, which remain stable over extended periods of time. It is conceivable that these high-density rings trap discoseismic modes in the inner disk. As proposed in \citet{Kato2004}, the resonant coupling between such p-modes and g-modes and a warp might lead to oscillations in the disk with a frequency substantially above the maximum radial-epicyclic frequency. In future work we plan to test this interesting hypothesis.

Another puzzling aspect is the detection of HFQPO pairs in a 3:2 frequency ratio in various sources (e.g. \citealp{Ingram_Motta2019}). In the literature these 3:2 frequency ratios were explained by the resonant coupling of the radial and vertical epicyclic and/or Keplerian frequencies at certain radii (e.g. \citealp{Kluzniak_Abramowicz2002, Abramowicz_Kluzniak_2001}) or discoseismic modes in a warp (e.g. \citealp{Kato2004}). However, there is no evidence for either a vertical epicyclic and/or Keplerian oscillation in the power spectra of radial mass flux, midplane density, and scaleheight of the disk in a 3:2 ratio (see Appendix \ref{sec:morePSDs}). Here we propose that such ratios may be explained by a corona, which sandwiches the accretion disk, but is not present in our simulations. Recent radiative GRMHD simulations have, for example, demonstrated that large scale magnetic fields can pressure support a geometrically thick corona-like structure around a very thin, radiation pressure supported, accretion disk \citep{LancovaAbarcaEtAl2019}. Since the corona is much thicker (e.g. \citealt{Liska_phase_lag}), it is conceivable that the center of the dense ring of gas associated with the corona extends to smaller radii away from the tearing radius. This might increase the coronal oscillation frequency. However, to explain a 3:2 frequency ratio between the disk and corona generated HFQPO, the corona would need to have a specific geometry that favours such ratios. This can be tested by the next-generation of radiative GRMHD simulations.

In addition, while LFQPOs are observed in various spectral states and evolve over a wide range of frequencies, HFQPOs are only seen in specific states at very specific frequencies that do not seem to evolve (e.g. \citealt{Ingram_Motta2019}). This implies, within the framework of our model, that a very narrow range of spectral states have favourable conditions for HFQPOs. Future radiative GRMHD will be required to test if changes in the disk geometry, such as the motion of the truncation radius, and/or the presence of large scale magnetic flux, will lead to damping and/or decoherence of the HFQPO in certain spectral states. Some evidence for this was presented in \citet{Motta_Belloni_Stella2014}, which suggested that broad lower-frequency features in the power spectrum of GRO-J1655 may eventually shift to higher frequencies and evolve into narrow HFQPO peaks.

\subsection{Misalignment in XRBs}
The QPO mechanism presented in this work falls under the class of models that are based on the misalignment between the angular momentum of binary system and that of the black hole \citep[e.g.][and references therein]{StellaVietri1998,Fragile_Mathews_Wilson2001,Ingram_Motta2019}. While it is conceivable that accretion disks in AGN are considerably misaligned due to the randomness of gas infall and binary black hole mergers \citep[e.g.][]{Volonteri_Madau_Quataert2005}, this is not entirely clear for XRBs. Namely, one would expect that the black hole and it's companion would have formed out of the same cloud of gas, and, that their spins further align with the orbit plane during the binary evolution.  However, a black hole in an XRB can become substantially misaligned due to natal supernova kicks \citep{Fragos_Tremmel_Rantsiou2010,Martin_Tout_Pringle2010,Jonker_Nelemans2004,SalvesenPokawanvit2020} or multi-body interactions \citep[e.g.][]{AntoniniRodriguezEtAl2018,FragioneKocsis2020}. Though such processes are not expected to reproduce the extreme misalignment angles assumed in this work, analytical arguments and SPH simulations do suggest that accretion disks, which are thinner than presented in this work, can tear at much smaller tilt angles (e.g. \citealt{NixonEtAl2012, NealonEtAl2015, Dogan_Nixon_King2018,RajEtAl2021}). Thus we expect that the results presented in this paper are applicable to a wide variety of systems. 

\section{Conclusions}
\label{sec:Conclusion}
In this work we have presented the first GRMHD simulation of a geometrically thin ($h/r = 0.02$), highly tilted ($65^\circ$) accretion disk around a rapidly spinning BH ($a = 0.9375$), which was threaded by a $\beta \sim 7$ toroidal magnetic field. The absence of any \citet{BlandfordZnajek1977} jet suggests that this simulation is primarily applicable to the soft-intermediate and high-soft states of BHXRBs (e.g. \citealt{Fender_Belloni2004_Gallo2004}). The unprecedented resolution of $N_r \times N_{\theta} \times N_{\phi} =13440 \times 4608 \times 8192$ cells allowed us to study the structure and dynamics of the disk with unprecedented accuracy.

We found that the (effective) differential Lense-Thirring torque exerted by the spinning BH on the disk causes the disk to tear $\sim 10$ times over $t \gtrsim 140,000 r_g/c$. During each tearing cycle a subdisk of size $r_t \sim 7.5-15 r_g$ forms and starts precessing. The precession of the inner sub-disk is well described as ``warped solid body'', with a precession frequency that is given by the nodal precession frequency $\nu_{\rm LT}(\xi r_0)$, where $r_0$ is the tearing radius and the $\xi$ parameter accounts for the moment of inertia in the disk. We find that $\xi\simeq0.75$ gives a good match with the measured precession frequencies of $\nu_{prec} \sim 2.5$ Hz. Since the precession frequency remains relatively stable during each tearing cycle, the modulation in flux received from this precessing sub-disk is expected to give rise to coherent LFQPO emission.

In addition, disk tearing drives a global $m=0$ radial epicyclic oscillation of a dense ring of gas at the tearing radius. This translates into two HFQPO signatures in the radial mass flux during tearing cycle E7: A fundamental peak with centroid frequency $\nu_0 \approx 55$ Hz and quality factor of $Q\simeq11$, alongside a weaker first harmonic overtone with centroid frequency $\nu_0 \approx 110$ Hz (once scaled to a black hole mass of $10M_{\odot}$). In addition, we find a second HFQPO peak with centroid frequency $\nu_0 \approx 69$ Hz and no harmonic during tearing cycle E10. In future work we will demonstrate that this radial-epicyclic oscillation indeed translates into an observable HFQPO signature using general relativistic ray-tracing (West, Liska, Musoke et al in prep), and put unique constraints on the viewing angles under which this HFQPO is visible.

The overall lifetime of each tearing cycle is delimited by the $\lesssim 2s$ accretion timescale of the inner sub-disk. This suggests that the observed low and high frequency QPO emission must be a summation of the quasi-periodic emission generated during separate tearing events and predicts similar coherence intervals for low and high frequency QPOs (e.g. \citealt{van_den_Eijnden_Ingram_Uttley2016}). Longer duration GRMHD simulations will be necessary to verify that both of these statements are accurate.

In this work we used a simplified prescription \citep{Noble_Krolik_Hawley2009} to artificially cool the disk to the targeted scale-height. This cooling function assumes that Coulomb collisions equilibrate the temperature between ions and electrons, and, thus that the disk can cool efficiently. We note that a more accurate treatment of the radiative processes alongside the inclusion of two-temperature thermodynamics might allow a self-consistent vertical disk structure to form, potentially leading to the generation of a hot (precessing) two-temperature inner torus surrounded by a cool, thin accretion disk (i.e. a 'Truncated' accretion disk, \citealp[e.g.][]{Esin_McClintock_1997ApJ, IngramEtAl2009}). Such a precessing torus could potentially emit hard, Comptonized, radiation. We plan to explore this interesting possibility in future work.

\section*{Acknowledgements}
An award of computer time was provided by the Innovative and Novel Computational Impact on Theory and Experiment (INCITE) program under award PHY129. This research used resources of the Oak Ridge Leadership Computing Facility, which is a DOE Office of Science User Facility supported under Contract DE-AC05-00OR22725. GM is supported by a Netherlands Research School for Astronomy (NOVA), Virtual Institute of Accretion (VIA) postdoctoral fellowship. ML was supported by the John Harvard Distinguished Science Fellowship. We acknowledge the SURFsara data processing facility SPIDER (Symbiotic Platform(s) for Interoperable Data Extraction and Redistribution) under award EINF 1120 for postprocessing of simulation data. A. I. acknowledges support from the Royal Society. GM, ML and OP thank Alexander Tchekhovskoy, Henric Krawczynski and Andrew West for insightful discussions.

This work has utilised the Stingray Python library for spectral timing: \href{https://doi.org/10.5281/zenodo.4881255}{DOI: 10.5281/zenodo.4881255} \citep{Bachetti_Huppenkothen_2021_stingray03}, LMFIT: Non-Linear Least-Square Minimization and Curve-Fitting for Python package \href{https://doi.org/10.5281/zenodo.11813}{DOI: 10.5281/zenodo.11813}, NumPy \citep{Harris2020_NumPy}, Matplotlib \citep{Hunter2007_Matplotlib} and SciPy \citep{Virtanen2020_SciPy}.
 
\section*{Data Availability}

Processed simulation data will be made available at \href{https://doi.org/DOI:10.5281/zenodo.5830266}{DOI:10.5281/zenodo.5830266}.



\bibliographystyle{mnras}
\bibliography{example,astroZ} 



\appendix

\section{QPOs in tearing cycle E10}
\label{sec:E10}
\begin{figure}
    \centering
    \includegraphics[clip,trim=0.0cm 0.0cm 0.0cm 0.0cm,width=7.3cm]{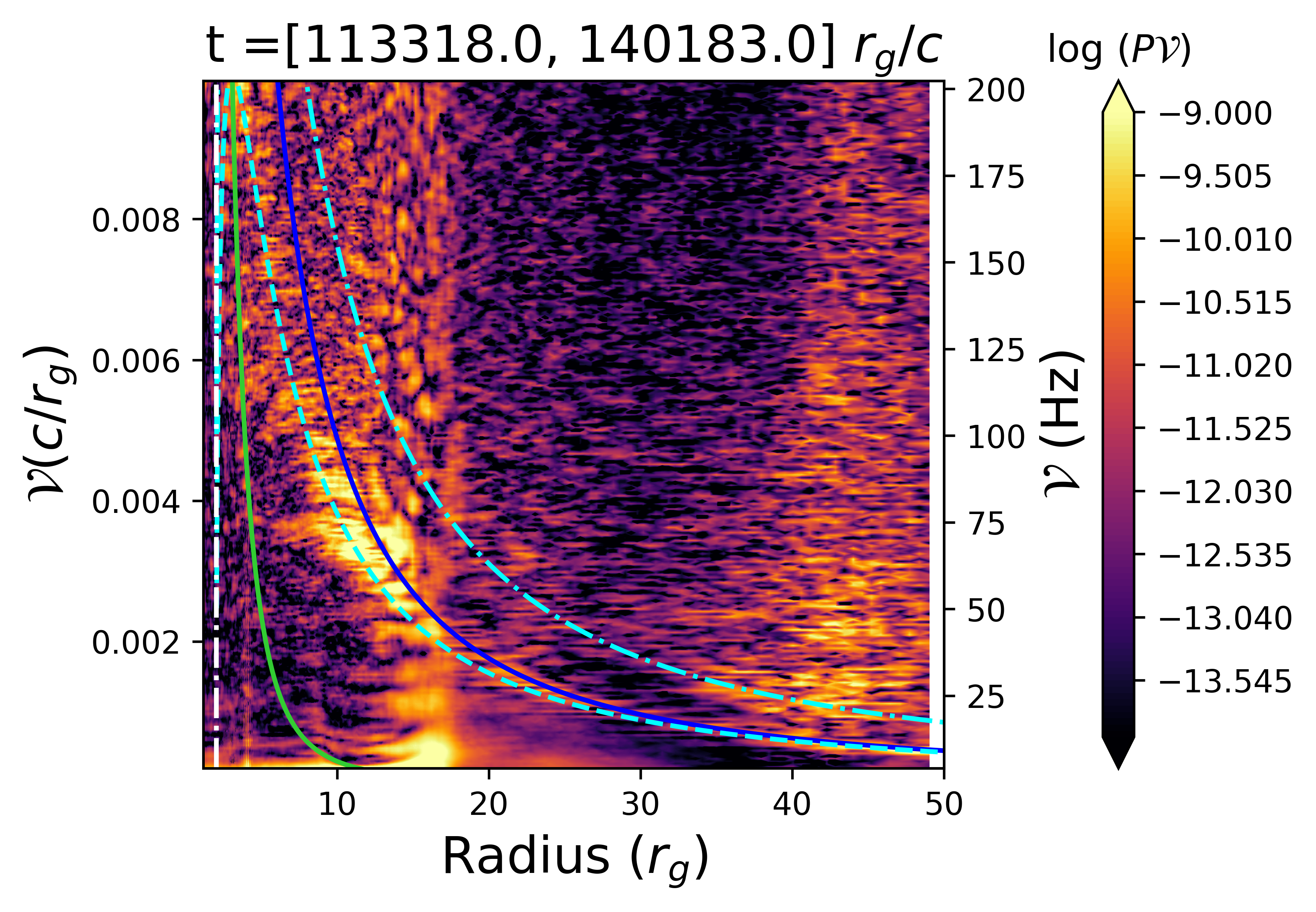}
    \includegraphics[clip,trim=0.0cm 0.0cm 0.0cm 0.85cm,width=7.3cm]{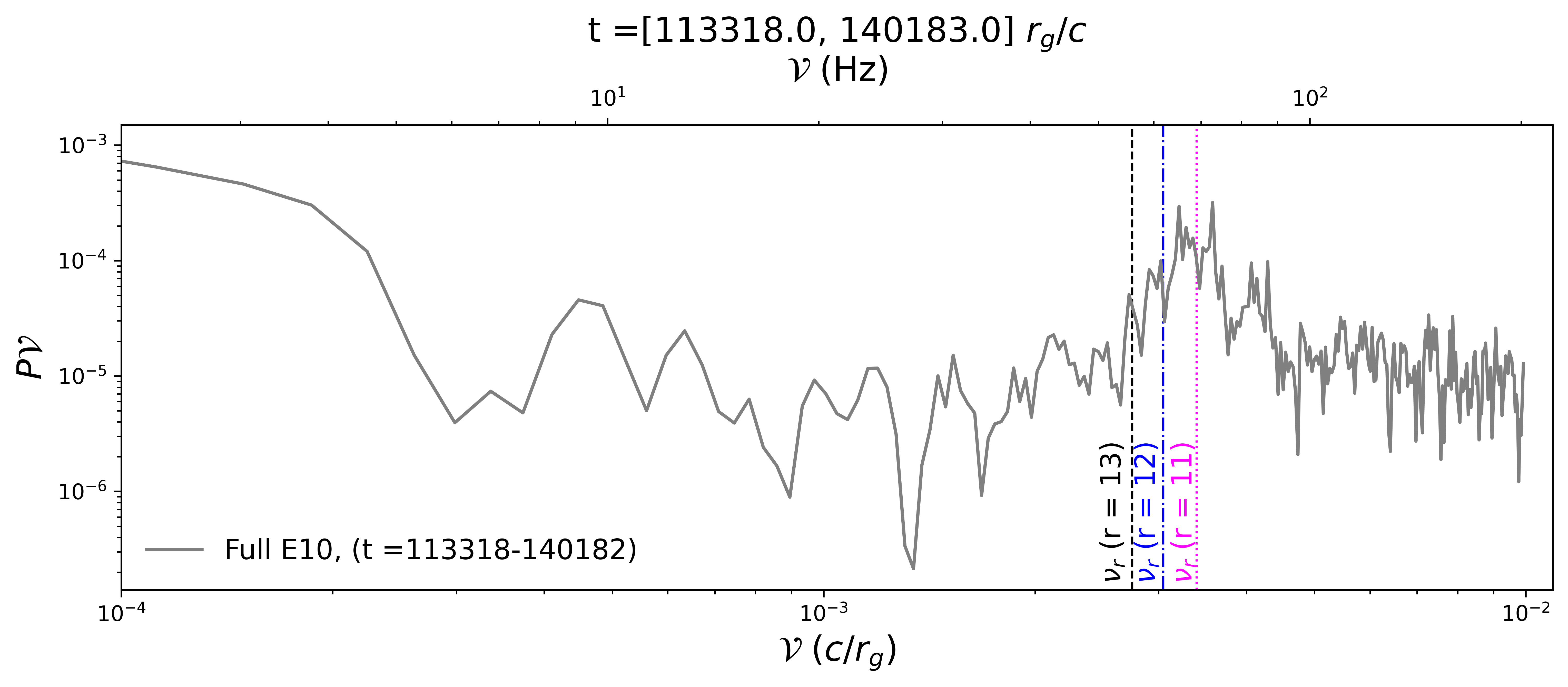}
    \caption{\textbf{Top:} PSD of the radial mass flux $\dot{M}$ for the full duration of the tearing cycle E10. The dashed cyan, solid blue and green curves correspond to the radial epicyclic, Keplerian and Lense-Thirring precession frequencies respectively. The colour bar shows log of the spectral power ($p \times \nu$). The dash-dot cyan curve corresponds to 2 times the radial epicyclic frequency. The vertical white dash-dot curve in each plot shows the location of the ISCO ($=2.044r_g$). The frequency $\nu$ in Hz (right $y$-axis in each plot) is computed for a 10 $M_\odot$ BH. \textbf{Bottom:} Power spectrum of the radial mass flux averaged over three discrete radii $11,12$ and $13 r_g$ and computed for the full duration of the tearing cycle E10. The lower $x$-axis shows the frequency in units of $c/r_g$ while the upper $x$-axis shows frequency in Hz computed for a 10 $M_\odot$ BH. The dashed black, blue and magenta curves show the radial epicyclic frequency $\nu_r$ at 13, 12 and 11 $r_g$ respectively.}
    \label{fig:powerspec_E10}
\end{figure}

In Figure \ref{fig:powerspec_E10} we show PSDs computed during tearing cycle E10 ($t = 113\,318- 140\,183 r_g/c$) listed in Table \ref{tab:disk_tearing}. In the top panel of Figure \ref{fig:powerspec_E10} we find a region of enhanced power on the radial epicyclic frequency curve (dashed cyan) at a location of $\sim12r_g$ with a frequency of $\sim 69$ Hz. This feature corresponds to a HFQPO peak with a centroid frequency $\nu_0 \approx 69$ Hz as shown on the power spectrum in the bottom panel of Figure \ref{fig:powerspec_E10} which shows the power spectrum averaged over three discrete radii $11,12$ and $13 r_g$. These radii correspond to locations at or near the position of the tearing radius during the time period in which the HFQPO is present in tearing cycle E10. The HFQPO peak in cycle E10 is broader and, since it is triggered at a smaller radius, has a higher frequency than the $55$ Hz HFQPO found in cycle E7 (Figure \ref{fig:multi-powerspec_BP}). During the time period in which the HFQPO is triggered in cycle E10 the tearing radius remains roughly fixed at $12 r_g$ and the centroid frequency of the peak roughly corresponds to the radial epicyclic frequency at $12 r_g$, as shown by the vertical dashed magenta curve in the bottom panel of Figure \ref{fig:powerspec_E10}. We note that the simulation ends before tearing cycle E10 is complete.

\section{Simultaneous low- and high-frequency QPOs and black hole spin}
\label{sec:measuring-spin}

In this paper we have argued that disk tearing can give rise to both LFQPO (due to relativistic precession of the inner sub-disk) and HFQPO (due to epicyclic motion at the edge of the inner sub-disk).  

\begin{figure}
    \centering
    \includegraphics[width=0.45\textwidth]{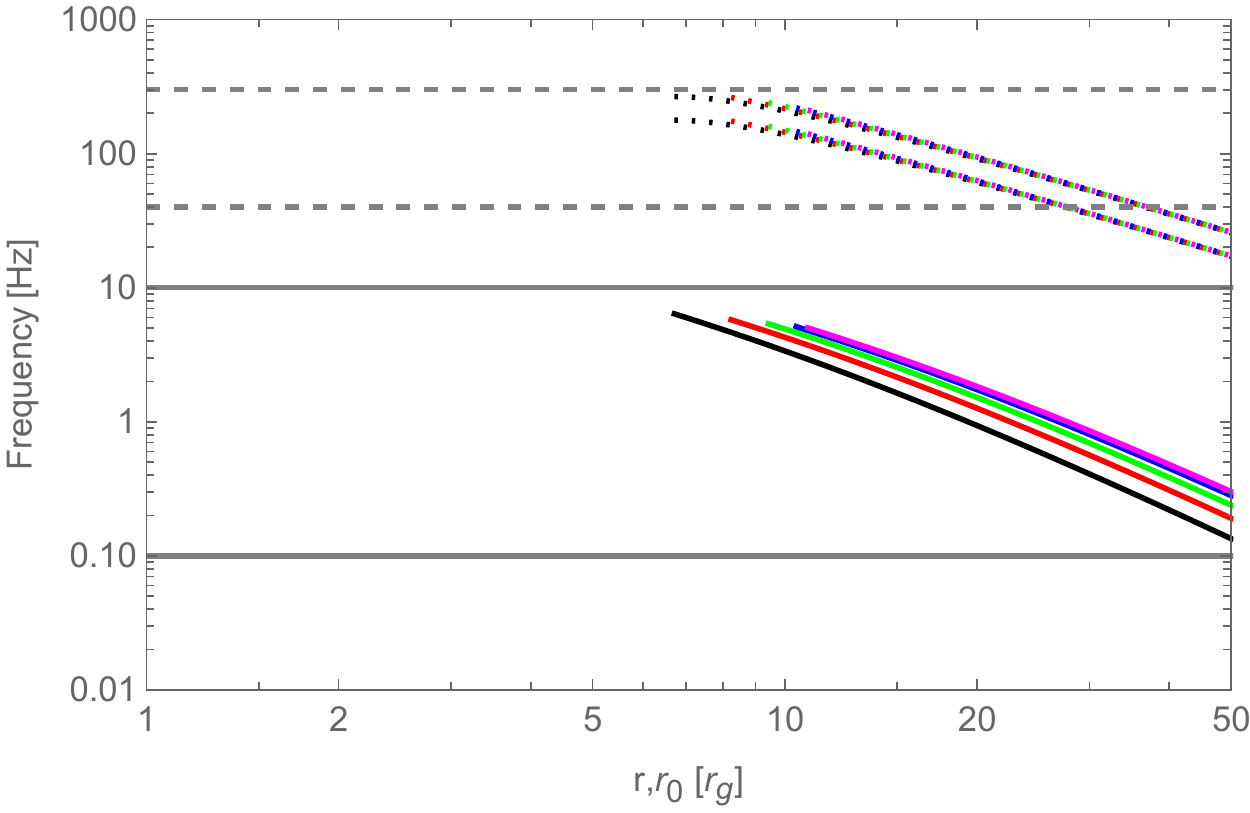}
    \caption{Simultaneous LFQPO and HFQPOs as a function of tearing radius. Here the LFQPO curves (solid lines) follows the relativistic precession model by \citet{IngramEtAl2009}  and the HFQPOs (dotted curves) are corresponding 2x and 3x overtones of the radial epicyclic frequency. We show spin values of $a \in \{0.3,0.5,0.7,0.9,0.998\}$ as black, red, green, blue and magenta lines respectively and assume a black hole mass of $10M_{\odot}$.}  
    \label{fig:LF_HF}
\end{figure}

For a given black hole mass, this implies a relation between the two types of QPOs that only depends on the angular momentum of the inner sub-disk and the black hole spin.  In Figure \ref{fig:LF_HF}, we show the frequencies for a range of black hole spin against the tearing radius.  Since the HFQPO often occur in pairs with frequency ratio 2:3, we give two HFQPO curves (dotted) corresponding to 2x and 3x the radial epicyclic frequency $\nu_{r}$.  
The precession frequency $\nu_{\rm prec}$ follows from the angular momentum distribution of the inner sub-disk which in principle can be extracted from the simulations.  For illustrative purposes, here we simply follow \cite{IngramEtAl2009} and model the disk with surface density $\Sigma = \Sigma_0 (r/r_i)^{-\xi}$, $h/r=0.2$, inner radius $r_{i}=(h/r)^{-4/5} a^{2/5}$ and adopt $\xi=0$ (see \citet{IngramEtAl2009} for a discussion of these parameters).  
The HFQPO depend very little on spin and range from $30-300\rm Hz$ for tearing radii between $r_i$ and $\simeq30 r_{g}$.  

\begin{figure}
    \centering
        \includegraphics[height=4cm]{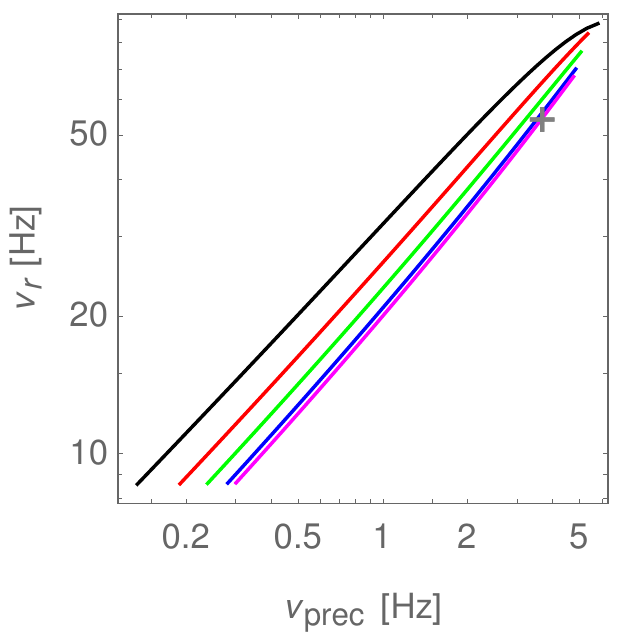}
        \includegraphics[height=3.93cm]{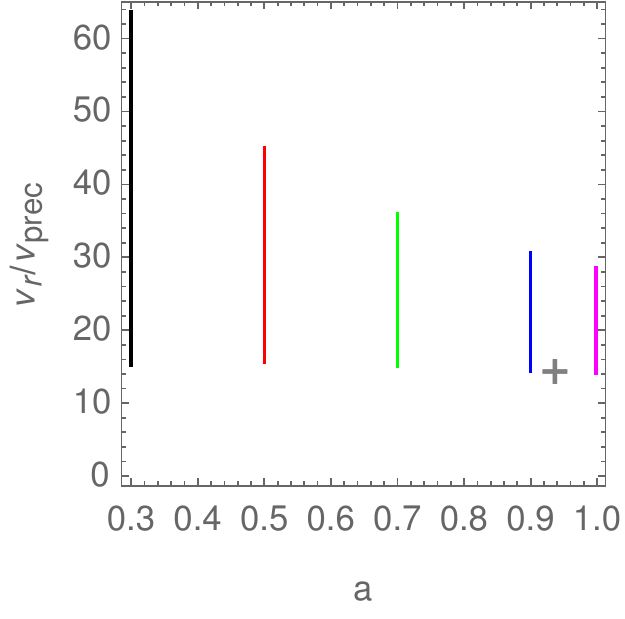}
    \caption{Radial epicyclic frequency against relativistic precession LFQPO frequency for the same parameters as in Figure \ref{fig:LF_HF} (left).  For a given black hole mass (here assumed $10M_{\odot}$), the spin is uniquely determined once both frequencies are observed.  In the right panel, we show the range of frequency ratios as function of the black hole spin, yielding a constraint on the spin for unknown black hole mass.  
    The gray cross shows the expectation for tearing at $13r_{\rm g}$ and spin $a=0.9375$ which matches well the obtained frequencies for tearing cycle E7.  }  
    \label{fig:epi_vs_LF}
\end{figure}

Under these assumptions, when the black hole mass is known, a simultaneous determination of $\nu_{r}$ and $\nu_{\rm prec}$ will uniquely determine the black hole spin as illustrated in the left panel of Figure \ref{fig:epi_vs_LF}. 
The constraint is however less strong if the black hole mass is unknown.  This is illustrated in the right panel of Figure \ref{fig:epi_vs_LF} which shows the adopted frequency \textit{ratios} as function of the black hole spin for tearing radii $r_0\in [r_i,50 r_{\rm g}]$.  It is evident from the figure that certain values of $\nu_{r}/\nu_{\rm prec}$ are incompatible with certain black hole spins. For example, $\nu_{r}/\nu_{\rm prec}>36$ rules out all spins \textit{above} $a=0.7$.  Setting a prior on black hole mass will further tighten this constraint for a given set of observed frequencies.  

\section{Ancillary PSDs for tearing cycle E7}
\label{sec:morePSDs}

\begin{figure}
    \centering
    \includegraphics[clip,trim=0.0cm 0.0cm 0.0cm 0.0cm,width=7.0cm]{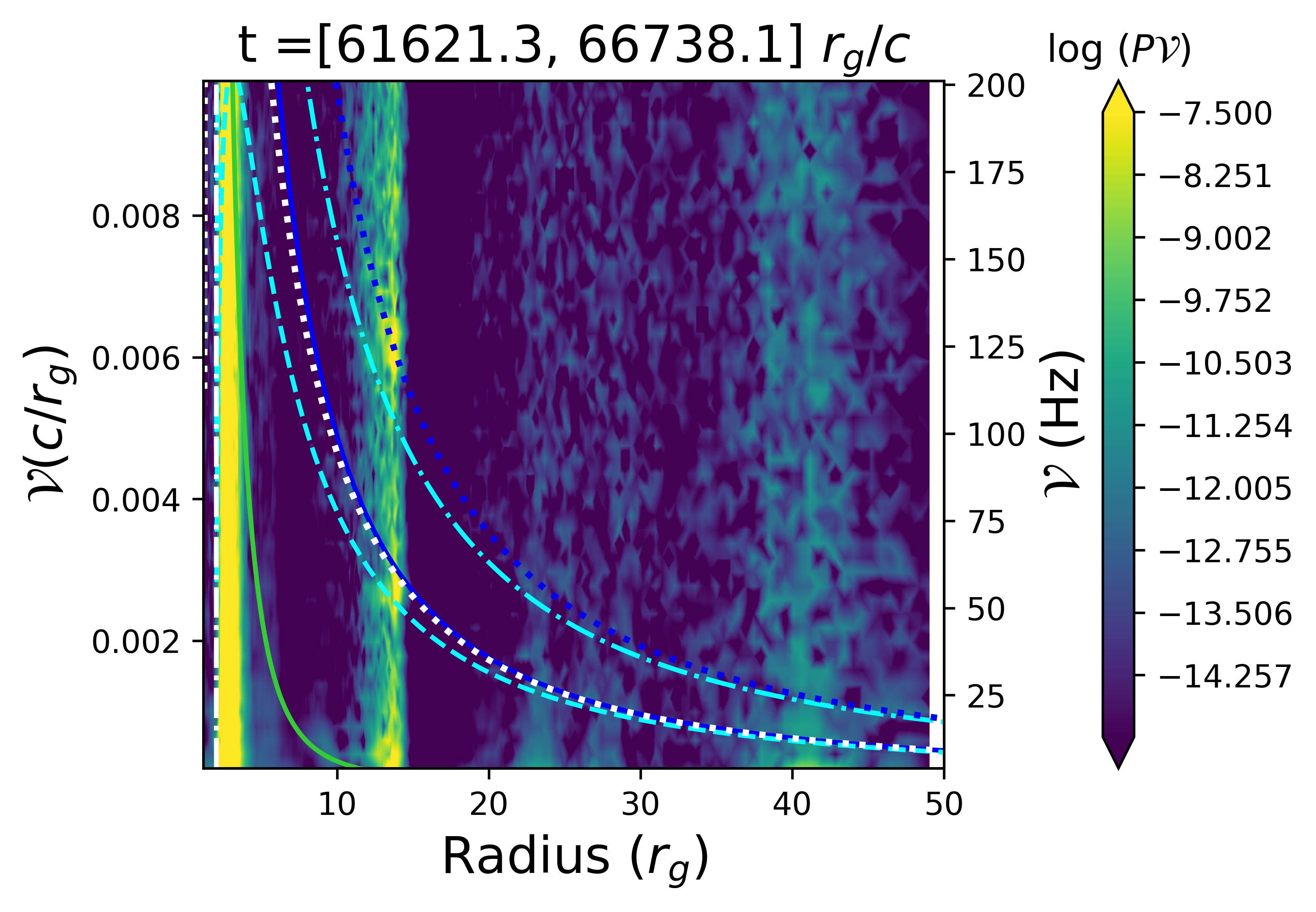}
    \includegraphics[clip,trim=0.0cm 0.0cm 0.0cm 0.0cm,width=7.0cm]{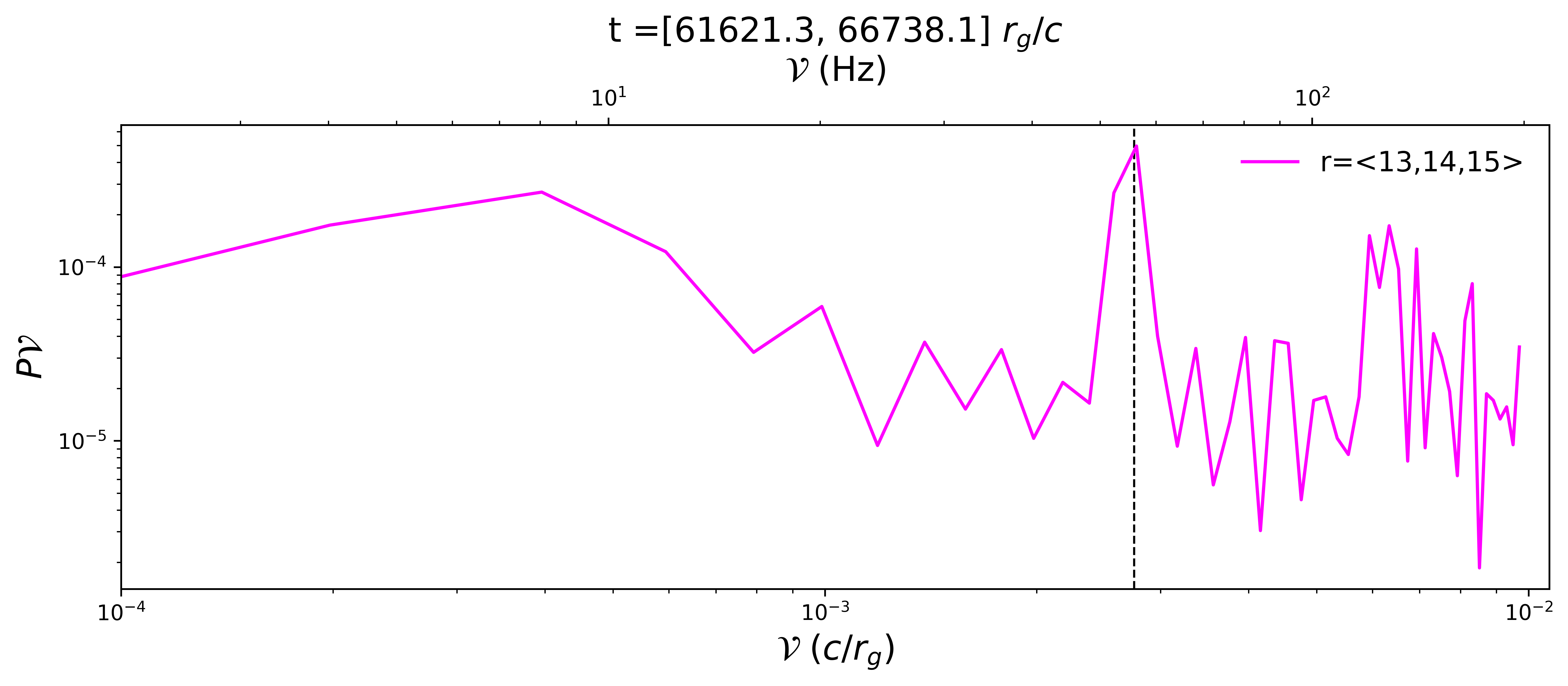}
    \caption{\textbf{Top:} PSD of $\bar{\rho}$ computed for the same time period as the top panel of Figure 4. Dotted blue curve is $2\times$ Keplerian frequency. Dotted white curve is vertical epicyclic frequency. Other curves are the same as in Figure \ref{fig:Mdot}. \textbf{Bottom:} Power spectrum computed for $\bar{\rho}$ and averaged over discrete radii of $13, 14$ and $15r_g$.}
    \label{fig:rho_avg}
\end{figure}

\begin{figure}
    \centering
    \includegraphics[clip,trim=0.0cm 0.0cm 0.0cm 0.0cm,width=7.0cm]{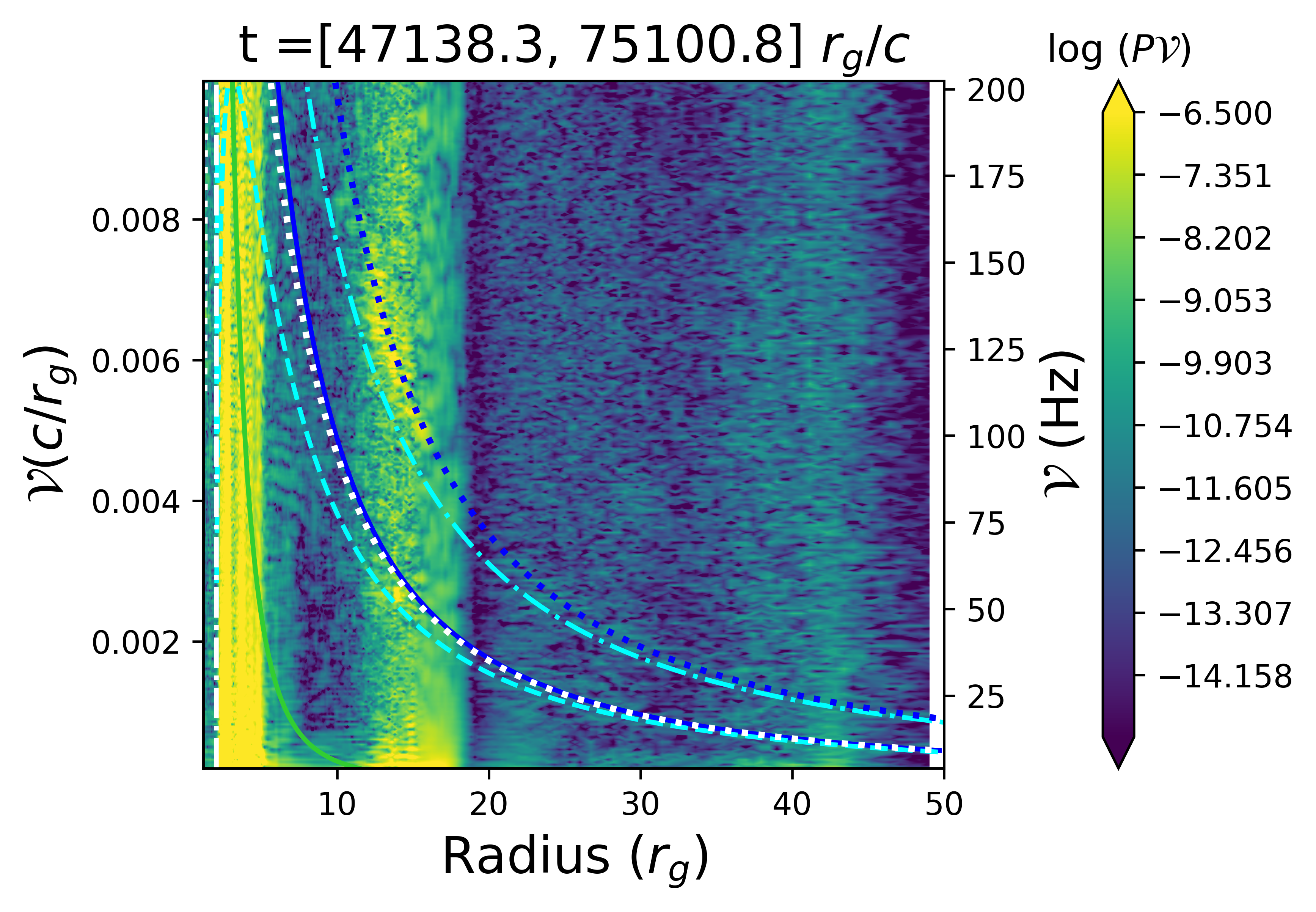}
    \includegraphics[clip,trim=0.0cm 0.1cm 0.0cm 0.0cm,width=7.0cm]{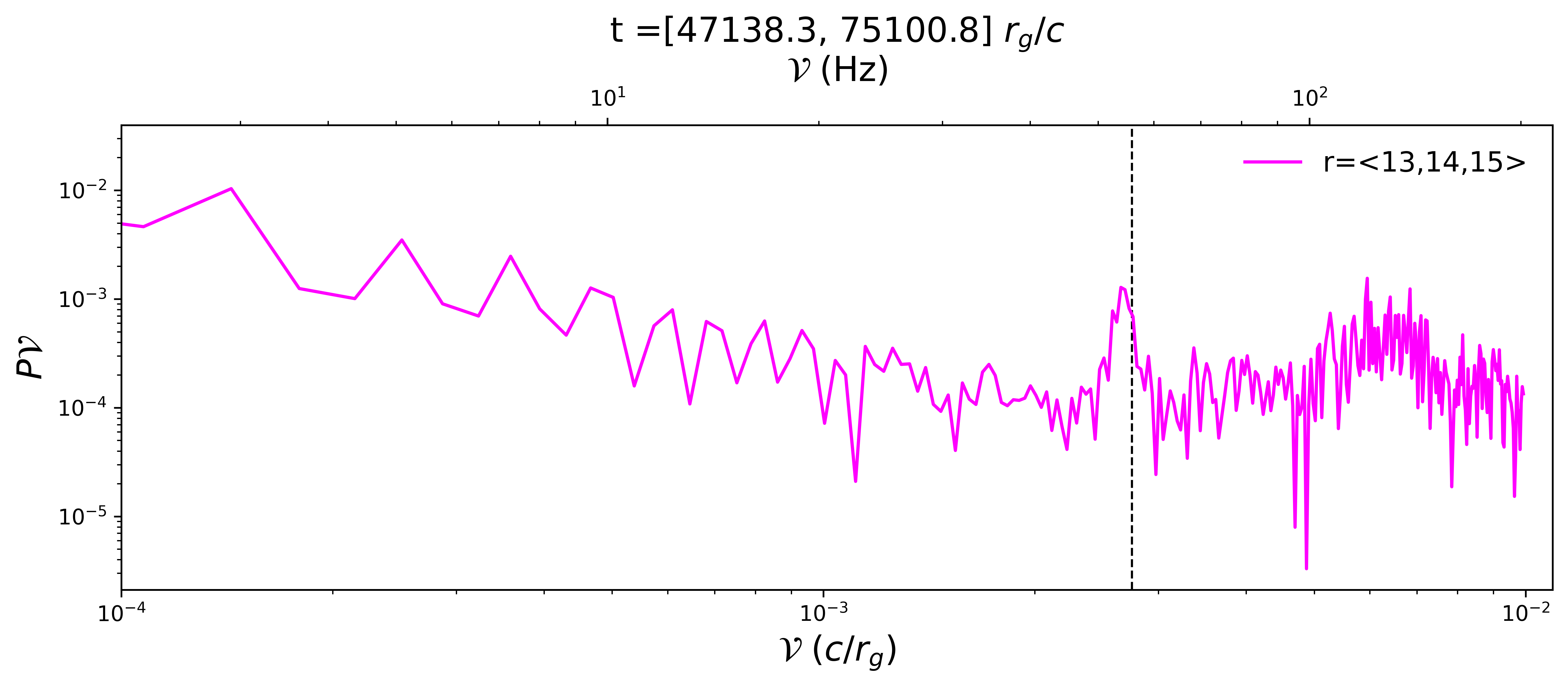}
    \caption{\textbf{Top:} PSD of $\bar{\rho}$ computed for the same time period as the middle panel of Figure 4. Dotted blue curve is $2\times$ Keplerian frequency. Dotted white curve is vertical epicyclic frequency. Other curves are the same as in Figure \ref{fig:Mdot}. \textbf{Bottom:} Power spectrum computed for $\bar{\rho}$ and averaged over discrete radii of $13, 14$ and $15r_g$.}
    \label{fig:rho_avg2}
\end{figure}

Here we present PSDs of complementary quantities for cycle E7. In principle, an inspection of the PSDs for different quantities (e.g. density, pressure and out-of-plane velocity) can help in identifying the (discoseismic-) modes, however such an investigation is beyond the scope of this work.    
Figures \ref{fig:rho_avg} and \ref{fig:rho_avg2} show PSDs of the averaged density for the same time interval as in the upper and middle panels of Figure \ref{fig:Mdot}. In addition to the characteristic frequency curves shown in Figure \ref{fig:Mdot} we also overplot the vertical epicyclic frequency \citep[e.g.][]{Nowak_Lehr1998} given by
\begin{equation}
    v_{\perp}^{2}  = \nu_{\Omega}^{2} \left( 1 - \frac{4a}{r^{3/2}} + \frac{3a^{2}}{r^{2}}\right) \text{ .}
\end{equation}

\begin{figure}
    \centering
    \includegraphics[clip,trim=0.0cm 0.0cm 0.0cm 0.0cm,width=7.0cm]{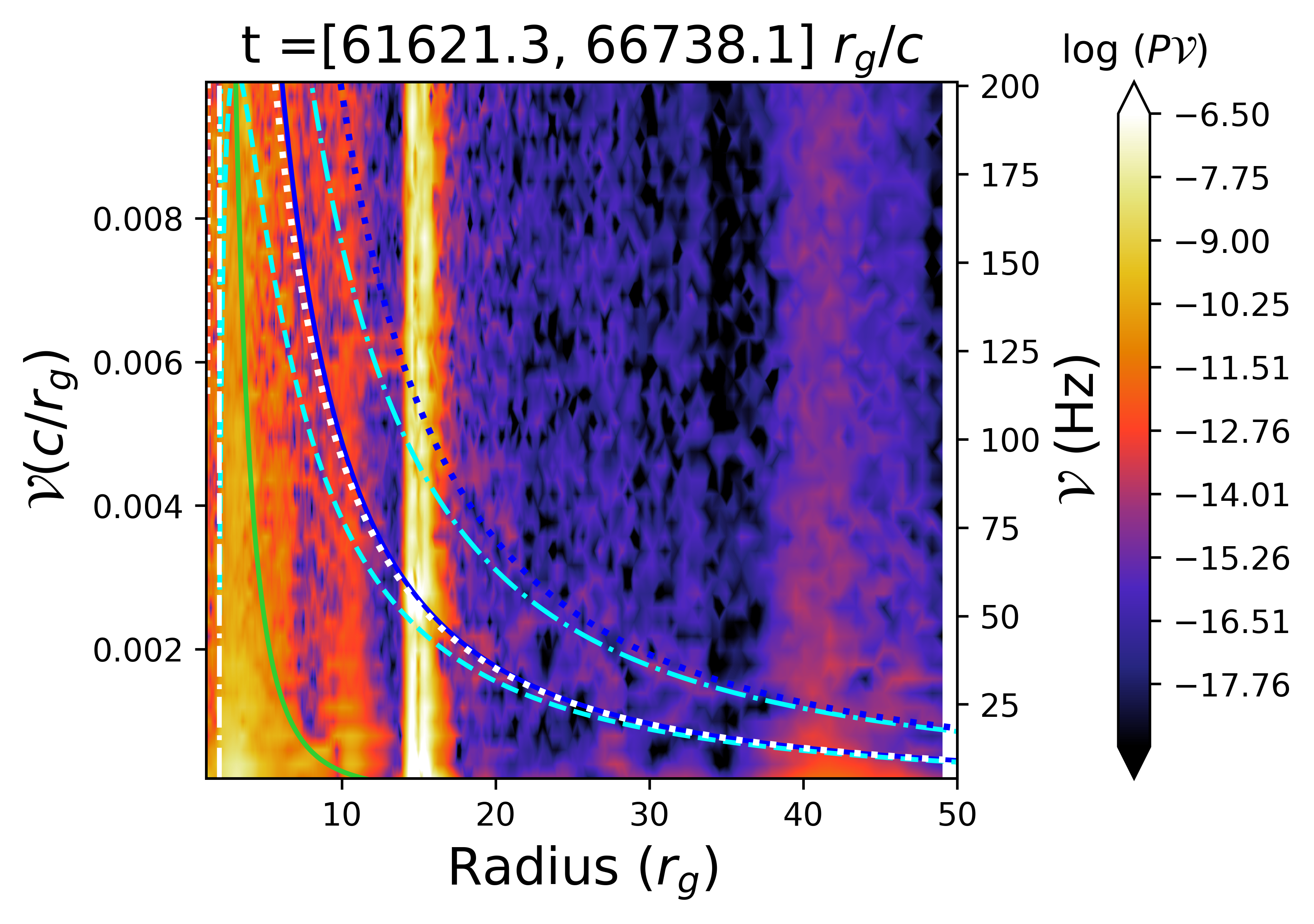}
    \includegraphics[clip,trim=0.0cm 0.0cm 0.0cm 0.0cm,width=7.0cm]{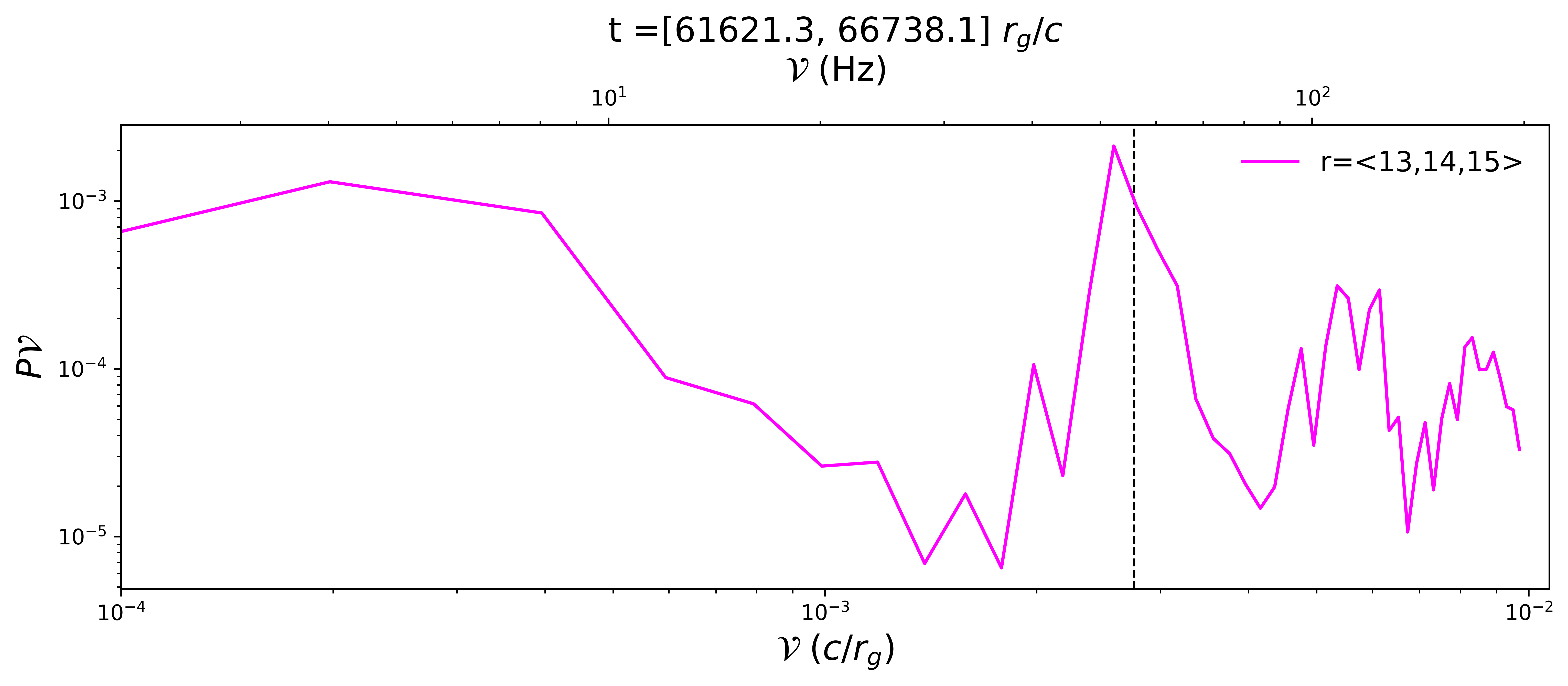}
    \caption{\textbf{Top:} PSD of scale height H/R computed for the same time period as the top panel of Figure 4. Dotted blue curve is $2\times$ Keplerian frequency. Dotted white curve is vertical epicyclic frequency. Other curves are the same as in Figure \ref{fig:Mdot}. \textbf{Bottom:} Power spectrum computed for H/R and averaged over discrete radii of $13, 14$ and $15r_g$.}
    \label{fig:hor}
\end{figure}

\begin{figure}
    \centering
    \includegraphics[clip,trim=0.0cm 0.0cm 0.0cm 0.0cm,width=7.0cm]{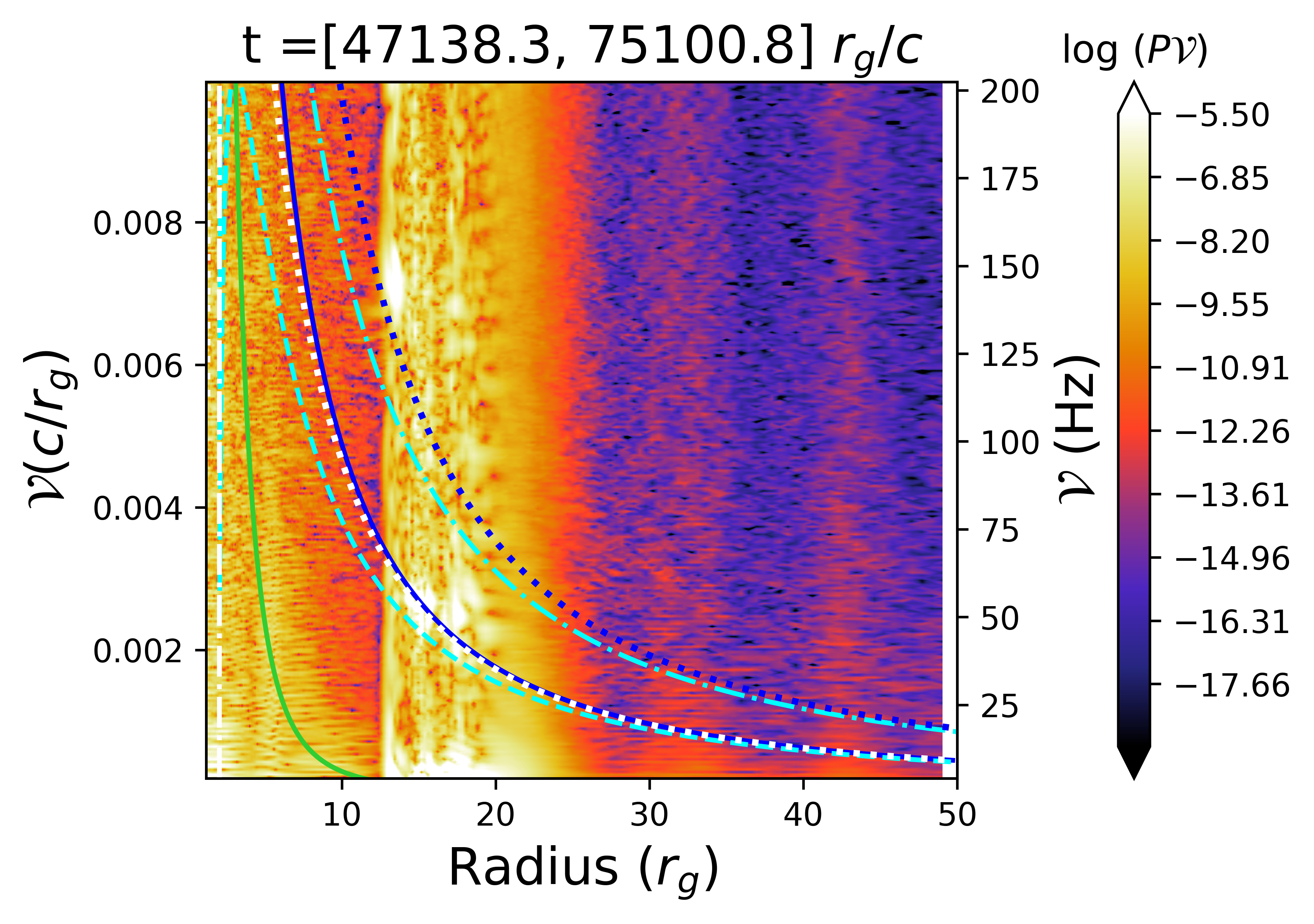}
    \includegraphics[clip,trim=0.0cm 0.0cm 0.0cm 0.0cm,width=7.0cm]{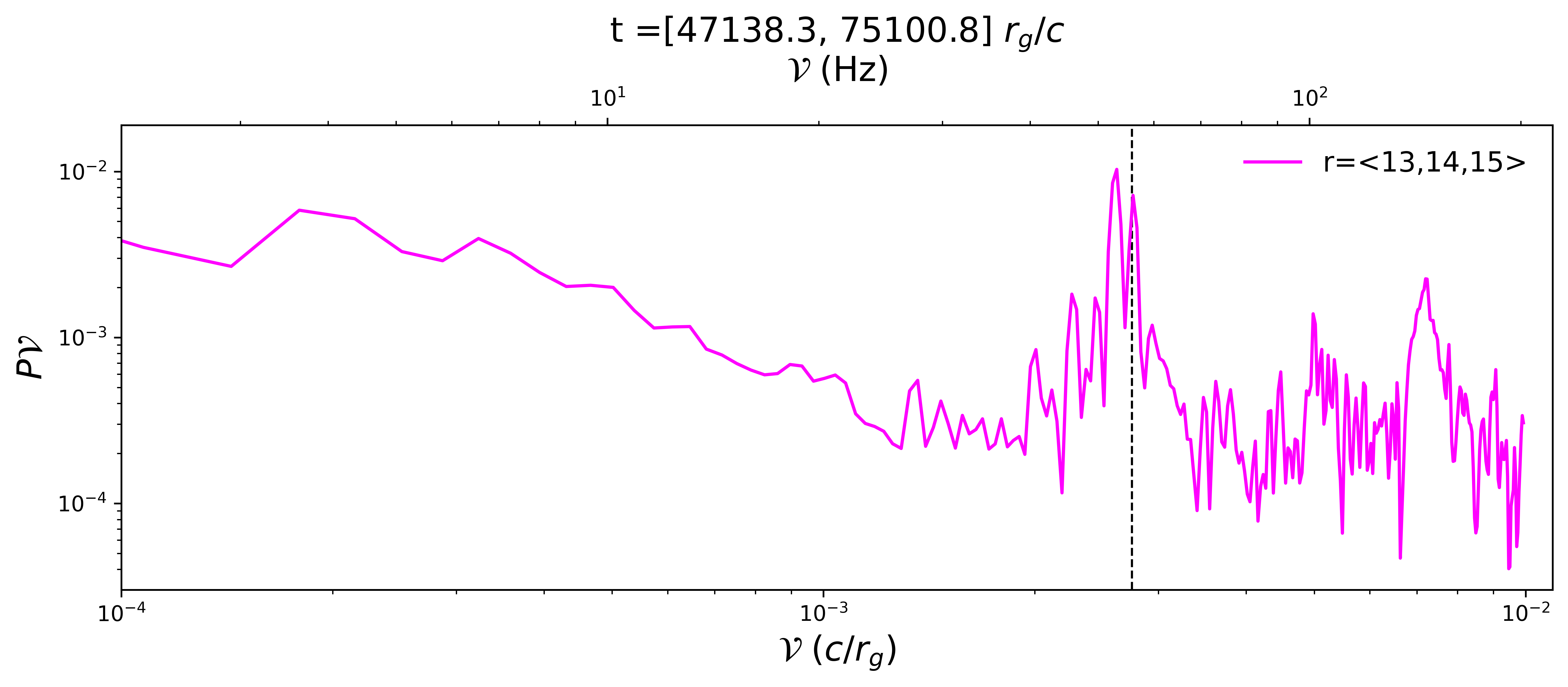}
    \caption{\textbf{Top:} PSD of scale height H/R computed for the same time period as the middle panel of Figure 4. Dotted blue curve is $2\times$ Keplerian frequency. Dotted white curve is vertical epicyclic frequency. Other curves are the same as in Figure \ref{fig:Mdot}. \textbf{Bottom:} Power spectrum computed for H/R and averaged over discrete radii of $13, 14$ and $15r_g$.}
    \label{fig:hor2}
\end{figure}

Figures \ref{fig:hor} and \ref{fig:hor2} give the corresponding PSDs of the scale height.  
Overall, the PSDs are similar to the PSD of $\dot{M}$ discussed in Section \ref{sec:HFQPO_detection}, most importantly the $55\,\rm Hz$ signal is visible in all quantities we have checked. However it is also apparent that the spectra shown here seem to have stronger harmonics at two times the Keplerian frequency. A more detailed investigation is left for future work.  

\bsp	
\label{lastpage}
\end{document}